\documentclass[10pt,aps,twocolumn,nofootinbib,superscriptaddress,notitlepage]{revtex4-1}

\usepackage{amssymb,amsmath}
\usepackage{graphicx}
\usepackage{color}
\usepackage{hyperref}
\usepackage{listings}
\usepackage{bm}
\usepackage{dsfont}

\usepackage{algorithm}
\usepackage{algpseudocode}
\usepackage{longtable}

\makeatletter
\gdef\@ptsize{0}
\let\@currsize\normalsize
\makeatother
\usepackage{setspace}
\hypersetup{
    unicode=false,          % non-Latin characters in Acrobat’s bookmarks
    pdftoolbar=true,        % show Acrobat’s toolbar?
    pdfmenubar=true,        % show Acrobat’s menu?
    pdffitwindow=false,     % window fit to page when opened
    pdfauthor={Ryan Gallagher},     % author
    colorlinks=true,       % false: boxed links; true: colored links
    linkcolor=blue,          % color of internal links
    citecolor=blue,        % color of links to bibliography
}
\graphicspath{{figures/}}

\makeatletter
\renewcommand{\fnum@figure}{\textbf{Figure \thefigure}}
\makeatother

\makeatletter
\renewcommand{\fnum@table}{\textbf{Table \thetable}}
\makeatother

\newcommand{\A}{\mathbf{A}}

\newcommand{\p}{\mathbf{p}}
\newcommand{\n}{\mathbf{n}}
\newcommand{\m}{\mathbf{m}}
\newcommand{\M}{\mathbf{M}}

\begin{document}

% ---------------------------------------------------------
% ---------------------- Author Info ----------------------
% ---------------------------------------------------------

\title{A Clarified Typology of Core-Periphery Structure in Networks}

\author{Ryan J. Gallagher}
\email{gallagher.r@northeastern.edu}
\affiliation{Network Science Institute, Northeastern University}
\author{Jean-Gabriel Young}
\affiliation{Center for the Study  of Complex Systems, University of Michigan}
\author{Brooke Foucault Welles}
\affiliation{Network Science Institute, Northeastern University}
\affiliation{Department of Communication Studies, Northeastern University}
\date{\today}

% ---------------------------------------------------------
% ----------------------- Abstract ------------------------
% ---------------------------------------------------------

\begin{abstract}
Core-periphery structure, the arrangement of a network into a dense core and sparse periphery, is a versatile descriptor of various social, biological, and technological networks.
In practice, different core-periphery algorithms are often applied interchangeably, despite the fact that they can yield inconsistent descriptions of core-periphery structure.
For example, two of the most widely used algorithms, the $k$-cores decomposition and the classic two-block model of Borgatti and Everett, extract fundamentally different structures:
the latter partitions a network into a binary hub-and-spoke layout, while the former divides it into a layered hierarchy.
We introduce a core-periphery typology to clarify these differences,
along with Bayesian stochastic block modeling techniques
to classify networks in accordance with this typology.
Empirically, we find a rich diversity of core-periphery structure among networks.
Through a detailed case study, we demonstrate the importance of acknowledging this diversity and situating networks within the core-periphery typology when conducting domain-specific analyses.
\end{abstract}

\maketitle

% ---------------------------------------------------------
% ----------------------- Main Text -----------------------
% ---------------------------------------------------------

\section{Introduction}

Core-periphery structure is a fundamental network pattern, referring to the presence of two qualitatively distinct components: a dense ``core'' of tightly connected nodes, and a sparse ``periphery'' of nodes loosely connected to the core and amongst each other. This pattern has helped explain a broad range of networked phenomena, including online amplification \cite{barbera2015critical}, cognitive learning processes \cite{bassett2013task}, technological infrastructure organization \cite{alvarez2008k,carmi2007model}, and critical disease-spreading conduits \cite{kitsak2010identification}. It applies so seamlessly across domains because it provides a succinct mesoscale description of a network’s organization around its core. By decomposing a network into core and peripheral nodes, core-periphery structure separates central processes from those on the margin, allowing us to more precisely classify the functional and dynamical roles of nodes with respect to their structural position. The analytic generality of this approach, together with the relative ubiquity of core-periphery structure among networks, makes core-periphery structure an indispensable methodological concept in the network science inventory.

Several methods and algorithms exist for extracting core-periphery structure from networks \cite{malliaros2020core}. They take on a variety of mathematical and algorithmic forms, ranging from statistical inference \cite{zhang2015identification,peixoto2017bayesian,kojaku2017finding}, spectral decomposition \cite{cucuringu2016detection,tudisco2019nonlinear}, and diffusion mapping \cite{della2013profiling} to motif counting \cite{ma2018detection}, geodesic tracing \cite{cucuringu2016detection}, and model averaging \cite{rombach2014core}. 
These algorithms exhibit a creative diversity of approaches for extracting core-periphery structure, and each is motivated by imagery of how core and peripheral nodes connect to one another. However, underneath the different high-level descriptions of each model, there are varying and often inconsistent assumptions about how the core and periphery are mutually connected, and how core-periphery structure is reflected in a network. As a result, despite the importance of core-periphery decomposition in answering substantive domain questions outside of network science, practitioners looking to apply these methods to their own fields are left without a warning that each algorithm has a different vision of what “core-periphery structure” actually means. This threatens the ability of researchers to draw valid conclusions about the structure and dynamics of numerous networks. By introducing a core-periphery typology that distinguishes between two qualitatively and quantitatively distinct structures, and by providing statistical techniques for determining where networks fall within that typology, we intend to make the distinction between various core-periphery structures clear, and enable reliable network inferences by scholars and practitioners.

\begin{figure*}[t]
\centering
\includegraphics[scale=.35]{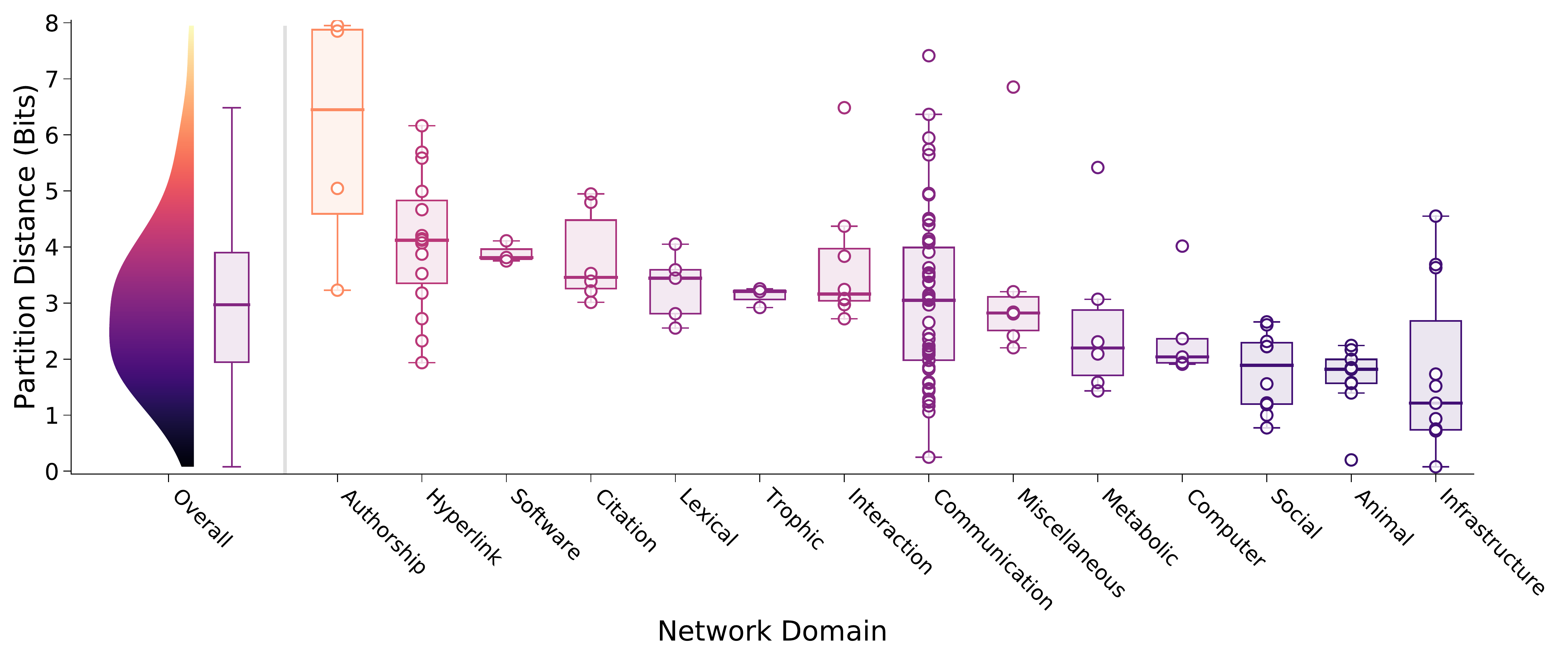}
\caption{Distributions by KONECT network domain \cite{kunegis2013konect} of distances between partitions extracted by the two-block model \cite{borgatti2000models} and the $k$-cores decomposition \cite{seidman1983network}. Distance is measured by the variation of information \cite{meilua2007comparing}, where higher values indicate more dissimilar partitions. Thick lines in each domain’s box plot indicate the median difference. Detailed results are reported in Appendix~\ref{appendix-konect}.}
\label{fig01-similarities}
\end{figure*}

The two types of core-periphery characterizations in our typology are well-exemplified by two of the most popular approaches for identifying core-periphery structure in networks. The first, which we refer to as the ``two-block model,'' is rooted in a definition originally proposed by Borgatti and Everett \cite{borgatti2000models}. Their mathematical formulation of core-periphery structure proposes that nodes are arranged into two groups, the core and the periphery, such that, ``core nodes are adjacent to other core nodes, core nodes are adjacent to some periphery nodes, and periphery nodes do not connect with other periphery nodes \cite{borgatti2000models}.'' This paints a \emph{hub-and-spoke} picture of core-periphery structure: there is a central hub of interwoven nodes, and a periphery that radiates out from that hub. The hub-and-spoke core-periphery formulation is at the backbone of network science methodology because it underlies many of the more sophisticated models that have been developed since Borgatti and Everett’s foundational work \cite{rombach2014core,zhang2015identification,kojaku2017finding}. Further, because the two-block formulation was originally proposed in the language of block models \cite{karrer2011stochastic}, it is often the \textit{de facto} statistical definition of core-periphery structure for many network scientists.

The second core-periphery characterization is reflected in the widely used $k$-cores decomposition. The $k$-core of a network is the largest subset of nodes in the network such that every node has at least $k$ connections. \cite{bollobas1998modern}. The $k$-cores define a hierarchy of $k$-shells, each of which consists of all the nodes in the $k$-core but not the $(k+1)$-core. The $k$-cores decomposition highlights a network's core-periphery structure by iteratively removing the $k$-shells \cite{seidman1983network}, starting with peripheral low-degree nodes in the outer shells and working towards embedded high-degree nodes in the inner cores. This algorithmic pruning process is accompanied by a suite of evocative language: the periphery is described as a series of ``shells,'' \cite{bollobas1998modern} ``onion layers,'' \cite{hebert2016multi} or ``leaves'' \cite{batagelj2003algorithm}, while the core is referred to as the ``epicenter,'' \cite{barbera2015critical} ``corona,'' \cite{goltsev2006k} or ``nucleus'' \cite{carmi2007model}. The language of the $k$-cores decomposition conjures up an image of a \emph{layered} core-periphery structure, composed of a nested sequence of layers that funnel towards a core. The scalable algorithm of the $k$-cores decomposition \cite{batagelj2003algorithm} has made it a practical tool for studying networks of all sizes, meaning that a number of applied network analyses implicitly assume a layered network arrangement.

By these accounts, it is clear that the layered and hub-and-spoke characterizations provide distinct descriptions of core-periphery structure. The differences between the hub-and-spoke and layered core-periphery characterizations are more than a linguistic sleight of hand though --- they can have repercussive consequences for substantive network analyses. In what follows, we first show that the structures extracted by the two most widely used algorithms --- the two-block model and the $k$-cores decomposition ---  diverge quantitatively across many empirical networks. To establish a statistically principled way of comparing these two classes of models, we then formulate both the hub-and-spoke and layered core-periphery structures as stochastic block models \cite{holland1983stochastic}, which allow us to encode the qualitative differences between the two characterizations and formulate an information-theoretic criterion of model fit \cite{peixoto2017bayesian,peixoto2017nonparametric}. With these tools, we analyze a suite of empirical networks and find a rich diversity of core-periphery structure that spans across the core-periphery typology. We finish with a case study of hashtag activism amplification, and emphasize how the choice of core-periphery model critically impacts the interpretation of substantive results. Our typology clarifies the distinct core-periphery structures that can emerge in networks, and provides a methodologically sound approach for disentangling those structures in practice.

\section{Results}

% ----------------------------------
\subsection{Inconsistent Core-Periphery Partitions}
\label{results-similarities}

We start by showing that the hub-and-spoke structure explicitly extracted by the two-block model \cite{borgatti2000models} and the layered structure implicitly suggested by the $k$-cores decomposition \cite{bollobas1998modern} provide fundamentally different descriptions of core-periphery structure for the same networks. To this end, we draw upon the Kolbenz Network Collection (KONECT) \cite{kunegis2013konect}, a diverse network repository that spans a number of social, biological, and technological domains. For each KONECT network, we extract the binary partition of nodes according to the two-block model \cite{lip2011fast} and the nested partition of nodes according to the $k$-cores decomposition \cite{batagelj2003algorithm}. We then measure the distance between these partitions via the variation of information (VI) \cite{meilua2007comparing,gates2019clusim}, and present the pairwise comparisons in Figure~\ref{fig01-similarities}. The VI is an information-theoretic measure, and is therefore expressed in bits per nodes. Intuitively, it can be thought of as the sum of information not shared by the two partitions. Hence, the more distant or dissimilar two partitions are, the larger the VI.

Across network domains, we find that the core-periphery partitions identified by the two-block model and $k$-cores decomposition are quite dissimilar, with an overall median VI of 2.9 bits per nodes. A normalized version of the VI, which can only be consistently interpreted for individual networks and not across data sets \cite{meilua2007comparing}, yields a median of 35\% the maximal value across domains. In other words, for each individual network, the partitions are about a third as distant as possible. Further, the differences in outcomes are not exhibited uniformly across domains. Some classes of networks (e.g. social, animal, infrastructure networks) see relatively more agreement between the two-block and $k$-cores partitions. For other network types (e.g. authorship, hyperlink, software networks), however, the two core-periphery algorithms almost always extract distant structures. Even within domains, there can be a wide heterogeneity in the similarities: for example, the range of distances is 7.2 bits for communication networks, 4.9 bits for infrastructure networks, and 4.2 bits for hyperlink networks. The results are similar when we use other measures to compare partitions, like the adjusted \cite{vinh2010information} or reduced \cite{newman2020improved} mutual information, and even when we match the partitions on sizes, to correct for the discrepancy between the number of $k$-cores and the number of groups in the two-block structure (see Appendix~\ref{appendix-konect}).

The results of Figure.~\ref{fig01-similarities} do not imply that one algorithm or the other is intrinsically flawed, but rather that the algorithms do not agree in general. If one's goal is only to describe a network, this disagreement is not an issue because each algorithm simply provides its own description of the network, whatever that may be \cite{young2018universality}. However, there is a strict \textit{statistical} sense in which the algorithms cannot both equally well-characterize a given network:  each core-periphery partition corresponds to a statistical description of the network, one of which will necessarily be more concise and precise than the other  \cite{olhede2014network,peixoto2013parsimonious}.
So, if we want to make network inferences based on core-periphery structure, we need to call on methods that can identify models that give better statistical descriptions of a network's structure than others.
Selecting models, however, requires that we have statistical models in the first place, and the notions of core-periphery structure that we have applied so far have only been defined through algorithms.
We therefore turn to Bayesian stochastic block models \cite{peixoto2017bayesian} to establish the missing connection between the two.

% ----------------------------------
\subsection{Core-Periphery Stochastic Block Models}
\label{results-models}

The stochastic block model is a general statistical model of a network's mesoscale structure \cite{holland1983stochastic}. At its core, it assumes that nodes belong to different groups, or ``blocks,'' such as the core and the periphery. These blocks then specify the probability that any two nodes are connected. More formally, suppose that we have the adjacency matrix $\A$ of an unweighted, undirected, simple network with $N$ nodes. We assume that there are a fixed number of $B$ blocks, and let the block assignments of all the nodes be recorded in $\theta$, a vector of length $N$ where $\theta_i = r$ indicates node $i$ belongs to block $r$. The probability that any two nodes in the network are connected is given by $\p$, a $B \times B$ matrix where $p_{rs}$ is the probability that a node in block $r$ connects to a node in block $s$. This is the defining characteristic of the stochastic block model: the block assignments completely determine the probability of connection between any two nodes.

\begin{figure*}[!t]
    \centering
    \includegraphics[scale=.7]{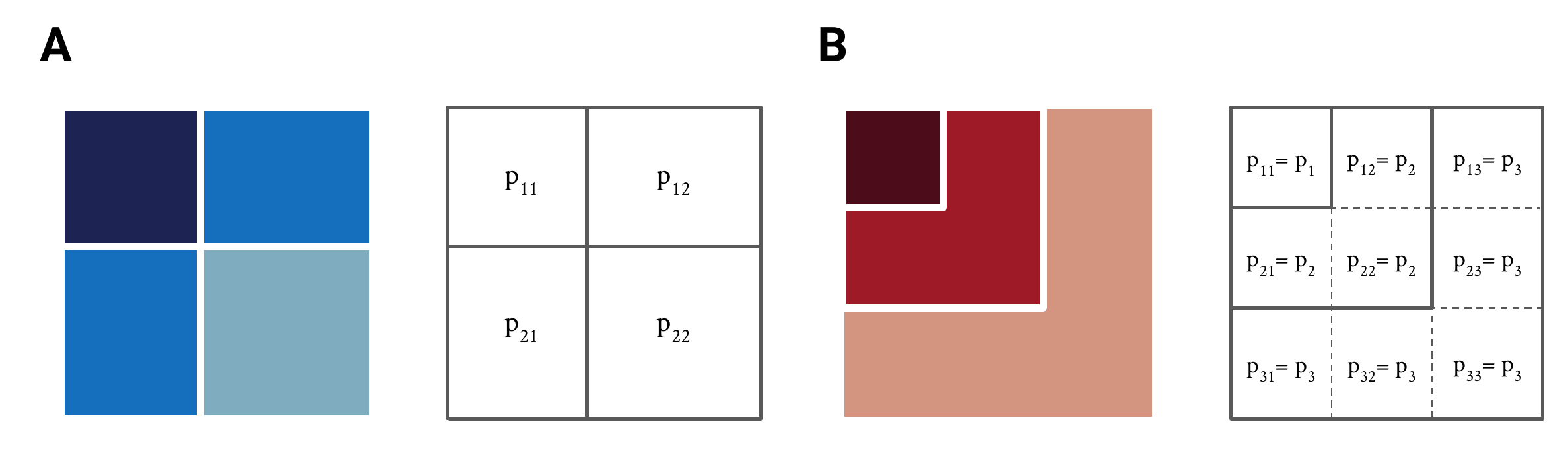}
    \caption{The core-periphery typology formalized through block model representations of the hub-and-spoke (\textbf{A}) and layered (\textbf{B}) structures. Each figure depicts the block connectivity matrix $\p$, where darker colors indicate higher densities of links. \textbf{(A)} The hub-and-spoke model is defined according to two blocks where $p_{11} > p_{12} > p_{22}$. \textbf{(B)} The layered core-periphery model is defined according to $\ell$ layers which are ordered as $p_1 > p_2 > \ldots > p_{\ell}$.}
    \label{fig-block-models}
\end{figure*}

In practice, we do not know the block assignments of the nodes $\theta$ or the probability of connection between blocks $\p$. We are interested, then, in the distribution $P(\theta, \p \mid \A)$, the probability we have a particular arrangement of nodes into blocks and connections between them, given our observed network data $\A$. Applying Bayes' rule, we have
\begin{equation}
    P(\theta, \p \mid \A)
        \propto
            P(\A \mid \theta, \p) P(\theta) P(\p).
\end{equation}

The posterior distribution $P(\theta, \p \mid \A)$ is proportional to three components: the  likelihood $P(\A \mid \theta, \p)$ of the network $\A$, the prior on the block assignments $P(\theta)$, and the prior on the block connectivity matrix $P(\p)$. We outline the standard setup of the likelihood and block assignment prior in the \hyperlink{methods-block-models}{Materials and Methods}. For constructing core-periphery stochastic block models, our main concern is with the prior on the block connectivity matrix. When we know that we want to model core-periphery structure, we only want to consider particular arrangements of connection probabilities $\p$, and that prior knowledge should be reflected in $P(\p)$.

We propose a core-periphery typology that contains two structures: the hub-and-spoke structure and the layered structure. Both characterizations can be phrased in the language of block models by arranging the block connectivities of $\p$ in different ways, as depicted in Figure~\ref{fig-block-models}. Through the Bayesian approach to the stochastic block modeling, we can alter the prior $P(\p)$ to encode these different arrangements, and constrain the model to adhere to those structures \cite{young2018universality}. The constrained models allow us to only consider networks with respect to the core-periphery typology and classify them appropriately according to the structure they exhibit.

The hub-and-spoke characterization specifies two blocks, one for the core and one for the periphery. If we let the core be denoted by the first block and the periphery by the second, then the original two-block model presented by Borgatti and Everett \cite{borgatti2000models} can be recovered by setting $p_{11} = 1, p_{12} = 1$, and $p_{22} = 0$. We consider a relaxation of this structure \cite{zhang2015identification}, which allows for flexibility in the connections by only requiring $p_{11} > p_{12} > p_{22}$. This configuration (shown in Figure~\ref{fig-block-models}A) conveys the intuition of the hub-and-spoke structure: there is a densely connected core moderately connected with a periphery which is only loosely connected amongst itself. Statistically, we enforce this constraint through a uniform prior over all block matrices that satisfy $0 < p_{22} < p_{12} < p_{11} < 1$. In notation, we write
\begin{equation}
    P(\p)  
        \propto 
            \mathds{1}_{\{0 < p_{22} < p_{12} < p_{11} < 1\}},
\end{equation}
where $\mathds{1}$ is the indicator function which takes on the value 1 if the constraint is satisfied, and 0 otherwise.

We can similarly formulate the layered block model. For convenience, we let $p_{rr} = p_r$, and assume there are $\ell$ layers, equal to the number of blocks $B$. To configure the layered structure shown in Figure~\ref{fig-block-models}B, we first specify
\begin{equation}
p_{rs} 
    = 
        p_{\max(r, s)},
\end{equation}
which binds the matrix $\p$ into layers. Like the hub-and-spoke model, we then order the layers through a uniform prior over all $\p$ that satisfy $0 < p_\ell < p_{\ell -1} < \ldots < p_1 < 1$,
\begin{equation}
    P(\p) 
        \propto 
            \mathds{1}_{\{0 < p_{\ell} < p_{\ell-1} < \ldots < p_{1} < 1\}}.
\end{equation}

All together, these constraints define the core-periphery stochastic block models. While the hub-and-spoke and layered formulations may seem innocuous, they complicate the analytical tractability of the standard stochastic block model significantly \cite{karrer2011stochastic,peixoto2017bayesian}. To infer the distributions of $\theta$ and $\p$ for the hub-and-spoke and layered block models, we therefore have to resort to numerical methods. The Gibbs sampling procedure that we devise is detailed in Appendix~\ref{appendix-estimation}.

\begin{figure*}[!t]
    \centering
    \includegraphics[scale=.3]{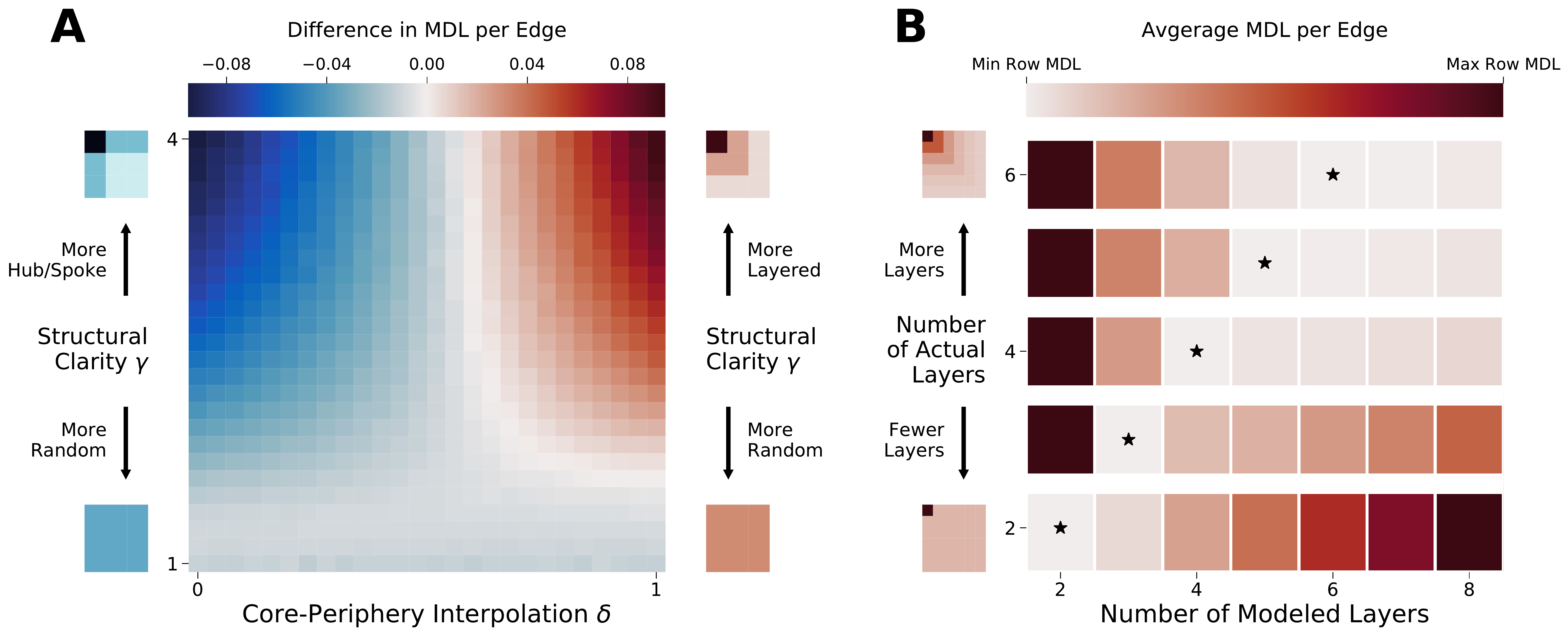}
    \caption{Inference results on synthetic network experiments. \textbf{(A)} The difference in MDL per edge between the layered model and hub-and-spoke model on networks constructed to have varying degrees of each structure. Negative values of difference in MDLs (blue) indicate that the hub-and-spoke model is a better model fit, while positive values (red) indicate that the layered model is a better fit. \textbf{(B)} The average MDL per edge of layered models on networks with planted layered structure. For each fixed number of actual layers in the synthetic networks (vertical axis), we run the inference with a varying number of modeled layers $\ell$ (horizontal axis). Stars ($\star$) indicates the number of modeled layers $\ell$ that yields the lowest MDL for a fixed number of planted layers.}
    \label{fig-synthetic-networks}
\end{figure*}

Recall that we have introduced these priors on the stochastic block model to determine the type of core-periphery structure that best describes a particular network. To make this determination, we first use the two models and associated inference techniques to find the core-periphery structures they each prescribe for that network. We then compare the inferred core-periphery structures to select the most statistically appropriate among the two. Formally, for a  partition $\theta_\mathcal{H}$ inferred through the hub-and-spoke model $\mathcal{H}$, and a partition $\theta_\mathcal{L}$ inferred through the layered model $\mathcal{L}$, we want to identify which model and its assignment of block labels to nodes is a better fit of the network data $\A$.
The answer is given by the the posterior odds ratio \cite{peixoto2017nonparametric},
\begin{equation}
\Lambda
    =
        \frac{P(\theta_\mathcal{H}, \mathcal{H} \mid \A)}
            {P(\theta_\mathcal{L}, \mathcal{L} \mid \A)}.
\end{equation} 
If the posterior odds ratio $\Lambda > 1$, then the hub-and-spoke model better characterizes the core-periphery structure of the network, while $\Lambda < 1$ implies that the layered model is a better descriptor. Assuming that we are agnostic about the models a priori, and so $P(\mathcal{H}) = P(\mathcal{L}) = 1/2$, we can equivalently consider
\begin{multline}
   - \log \Lambda
        =
            \Sigma_\mathcal{H} - \Sigma_\mathcal{L} 
       \\ =
            - \log P(\A, \theta_\mathcal{H} \mid \mathcal{H}) 
            +  \log P(\A, \theta_\mathcal{L} \mid \mathcal{L}),
\end{multline}
the difference between description lengths of the hub-and-spoke and layered models. The description length, $\Sigma_\mathcal{M} = -\log P(\A, \theta_\mathcal{M} \mid \mathcal{M})$, of a model $\mathcal{M}$ describes how well that model can compress the information expressed by a network's structure \cite{mackay2003information,peixoto2017nonparametric}. A model that is able to efficiently describe a network with a smaller number of parameters is a better descriptor of the network and will have a minimal description length. So, if the hub-and-spoke model $\mathcal{H}$ is a better descriptor of a network's core-periphery structure than the layered model $\mathcal{L}$, we will have a posterior odds ratio where $\Lambda > 1$ and, equivalently, a negative difference in description lengths, $\Sigma_\mathcal{H} - \Sigma_\mathcal{L} < 0$. We use the description length to quantify model fit, by considering either the pairwise difference in description lengths between two models or the minimum description length (MDL) across many models (see Appendix~\ref{appendix-model-fit} for numerical details). This measure allows us to distinguish which block model most aptly describes a particular network and properly situate it within the core-periphery typology.

We briefly note two connections of our core-periphery block models to prior work. With respect to the hub-and-spoke model, Zhang, Martin, and Newman \cite{zhang2015identification} identified the ordering $p_{11} > p_{12} > p_{22}$ as a relaxed version of the hub-and-spoke block structure introduced by Borgatti and Everett. However, they did not formally encode this constraint in their model and, instead, relied on the susceptibility of stochastic block models to heterogeneous degree distributions \cite{karrer2011stochastic} to retrieve core-periphery structure. With the respect to the layered model, Borgatti and Everett \cite{borgatti2000models} presented a special case of our model where $\ell = 2$, $p_1 = 1$ and $p_2 = 0$. However, given that those binary layer densities imply a network which consists of a connected core component surrounded by a cloud of isolate periphery nodes, they only briefly remarked on the model's limited conceptual utility. This is likely why a more general layered block model did not gain traction in later work.

% ----------------------------------
\subsection{Synthetic Network Experiments}
\label{results-synthetic-experiments}

As an essential validation step, we experimentally verify that our two models of core-periphery properly recover these  structures, when we know that they exists within a network. Our first experiment measures the capacity for the block models to discern between hub-and-spoke and layered structures. We generate synthetic networks according to the stochastic block model, and design $\p$ to have a known, ground truth core-periphery block arrangement. We configure $\p$ according to a two-parameter model (see \hyperlink{methods-3layer}{Materials and Methods} for details). The first parameter $\delta$ interpolates between hub-and-spoke and layered core-periphery structure: when $\delta =0$ the network has a known hub-and-spoke core-periphery structure, and when $\delta = 1$ the network has a known layered structure consisting of 3 layers. The second parameter $\gamma$ defines the structural clarity: when $\gamma = 1$ the network is random and neither model should be able to infer structure, and for large $\gamma$ the networks have a well-defined core-periphery structure.

\begin{figure*}[!t]
    \centering
    \includegraphics[scale=.43]{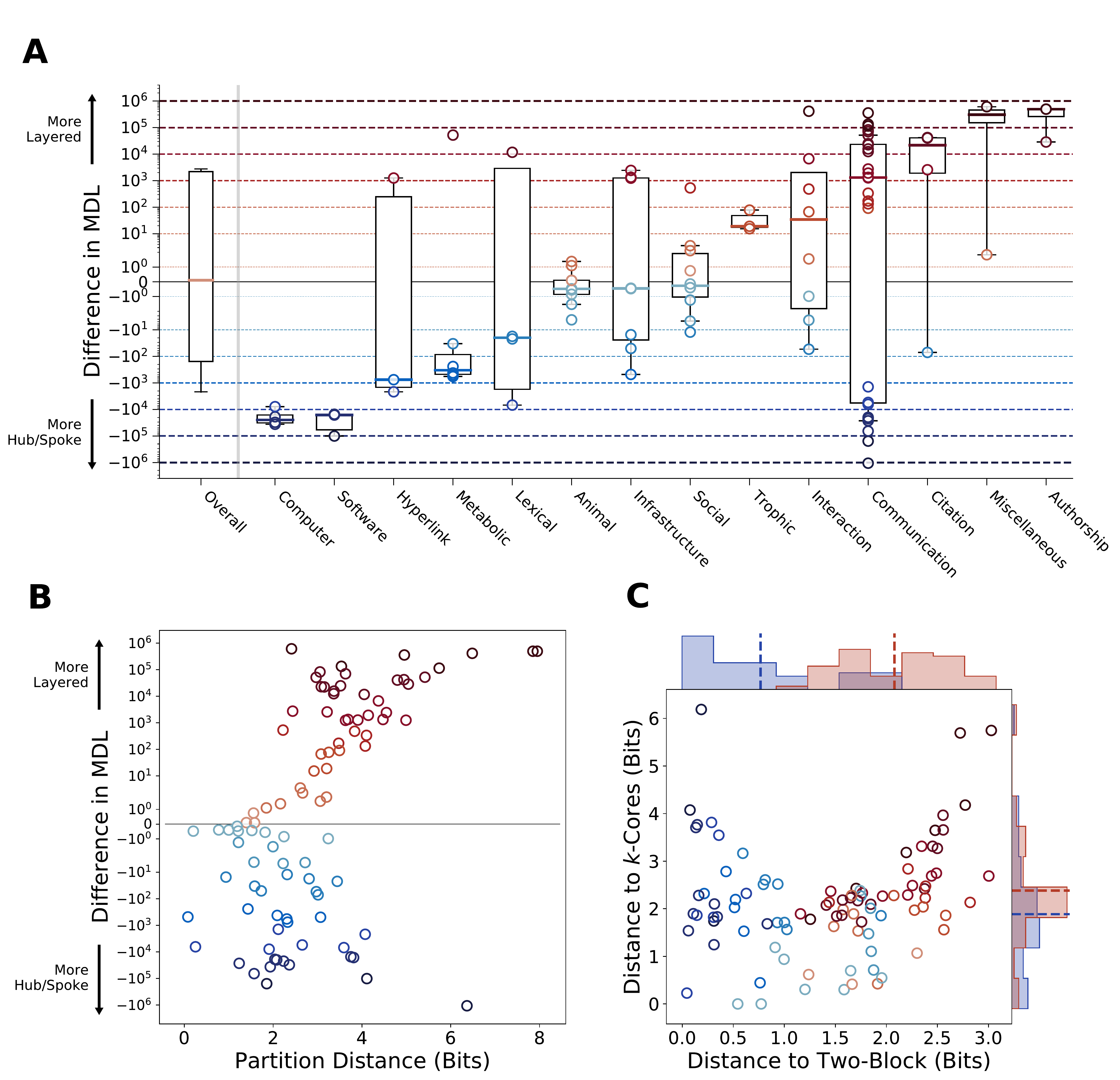}
    \caption{Structural diversity across the core-periphery typology. \textbf{(A)} Distributions of differences in minimum description length (MDL) between the best fit hub-and-spoke and layered models,  by network domain. Thick colored lines in each domain’s box plot indicate the median difference. \textbf{(B)} Difference in MDL plotted against the distance between the two-block and $k$-cores partitions for each KONECT network. \textbf{(C)} Distance between the best fit block model (either hub-and-spoke, indicated by blue, or layered, indicated by red) and the two-block and $k$-cores partitions. Histograms show the marginal distributions of distances, where dashed lines indicate the mean distance. Partition distance in all subplots is measured in bits according to the variation of information.}
    \label{fig-diversity}
\end{figure*}

The results, shown in Figure~\ref{fig-synthetic-networks}A, demonstrate that the block models effectively discern the two types of planted core-periphery structures. Within each regime of the interpolation parameter $\delta$, the minimum description length (MDL) appropriately identifies the correct model as a better description of the network structure, and it is more confident as the parameter reaches the boundaries of its range, at which point only that structure definitively exists in the network. For low values of structural clarity $\gamma$, for which the networks are purely random, the models have approximately equal minimum description lengths, and neither is strongly designated better in terms of model fit.

Our second experiment tests the layered model's ability to identify the appropriate number of layers in a synthetic network with a known layered structure. We start with a synthetic network of 6 equally sized layers, and progressively reduce the effective number of layers by merging layers until there are only 2 layers (see \hyperlink{methods-merged}{Materials and Methods} for details). For each fixed number of layers in the synthetic network, we run multiple layered core-periphery models for different choices of the parameter $\ell$, which designates the number of layers to infer. The results, given in Figure~\ref{fig-synthetic-networks}B, indicate that the average MDL accurately identifies the number of layers that exist in each synthetic network. This demonstrates that not only we can use the MDL for model selection between the hub-and-spoke and layered models, per the results of the first experiment, but we can also use it for choosing the number of layers.

% ----------------------------------
\subsection{Core-Periphery Diversity of Empirical Networks}
\label{results-diversity}

Having validated the core-periphery block models and the use of the minimum description length (MDL) as a measure of model fit, we establish the diversity of core-periphery structure expressed by empirical networks. For all networks with up to 200,000 nodes in the KONECT data set \cite{kunegis2013konect}, we infer partitions according to each of the hub-and-spoke and layered core-periphery models (see \hyperlink{methods}{Materials and Methods} for details). 

In Figure~\ref{fig-diversity}A, we show the breadth of network structure exhibited across the core-periphery typology.
As we clearly see, both the hub-and-spoke and layered core-periphery structures are expressed to a wide degree of intensity across all types of networks---neither model is a universal, best descriptor of core-periphery structure. Some classes of networks seem to be generally well-described by either just the hub-and-spoke characterization or just the layered characterization, but many more show a range of structure across the core-periphery spectrum. Communication networks in KONECT, for example, exhibit the full range of core-periphery prevalence across both characterizations. The diversity of core-periphery structure in these empirical networks demonstrates the danger in assuming a core-periphery type a priori, and the need to situate a network within the core-periphery typology to mitigate later downstream network mischaracterizations.

We also observe that a smaller portion of networks do not strongly exhibit either a hub-and-spoke or layered structure. In Figure~\ref{fig-diversity}B, we show that these networks are exactly those for which the two-block model and $k$-cores decomposition extract similar partitions. Further, the networks that have the most distant two-block and $k$-cores core-periphery partitions are also those that mostly strongly exhibit a hub-and-spoke or layered structure according to the block model description length. Figure~\ref{fig-diversity}C complements these findings by comparing the distances between the inferred partitions according to the stochastic block models and the partitions of the two-block model and $k$-cores decomposition. The hub-and-spoke partitions found with the SBM are consistently closer to the the two-block partitions than the k-core decomposition. The relationship between the layered partitions and the $k$-cores partitions is less sharp, with layered and hub-and-spoke partitions being about equally distant from the $k$-cores partitions on average. These results provide evidence that the two-block model is representative of the hub-and-spoke characterization. The layered characterization, however, finds partitions that are not quite the same as the k-core algorithm, due in part to the fact that it aggregates nodes in fewer layers.

% ----------------------------------
\subsection{Case Study: Hashtag Activism Amplification}
\label{results-case-study}

To emphasize the importance of distinguishing between hub-and-spoke and layered core-periphery structures, we briefly conclude with a case study of hashtag activism amplification. Social media are notable for creating spaces where historically disenfranchised individuals can come together and share their stories at an unprecedented scale \cite{benkler2006wealth,jackson2020hashtagactivism,papacharissi2015affective}. Hashtag activism, in particular, has been a critical vehicle for driving those marginalized voices into the mainstream public sphere \cite{jackson2020hashtagactivism,papacharissi2015affective}, as exemplified by hashtags like \#BlackLivesMatter and \#MeToo \cite{gallagher2019reclaiming,jackson2020hashtagactivism}. The amplification of those voices is a fundamentally networked process---the core consists of those who are most visible exactly because many peripheral amplifiers share the core's posts through emergent crowdsourcing \cite{barbera2015critical}. Core-periphery structure is a natural network model for such amplification processes.

Although those at the periphery of hashtag activism events are sometimes derided as ``slacktivists,’’ Barber\'a et al. \cite{barbera2015critical} demonstrated that the periphery contributes significantly to the amplification of core protest voices. We perform a similar analysis on the retweet network of the hashtag \#MeToo, a hashtag that highlighted the pervasiveness of sexual violence against women by creating a space for them to publicly disclose their experiences \cite{gallagher2019reclaiming,jackson2020hashtagactivism}. We fit hub-and-spoke and layered core-periphery models to the \#MeToo network and calculate the \emph{coreness} of each individual according to each model. The coreness, which varies from 0 to 1, indicates whether an individual is more likely to be situated in the periphery or core, respectively (see \hyperlink{methods-metoo}{Materials and Methods} for details). In line with prior work \cite{barbera2015critical}, we operationalize amplification by measuring the total \emph{reach} as the sum of the number of followers each individual in the network has.

\begin{figure}[!t]
    \centering
    \includegraphics[scale=.37]{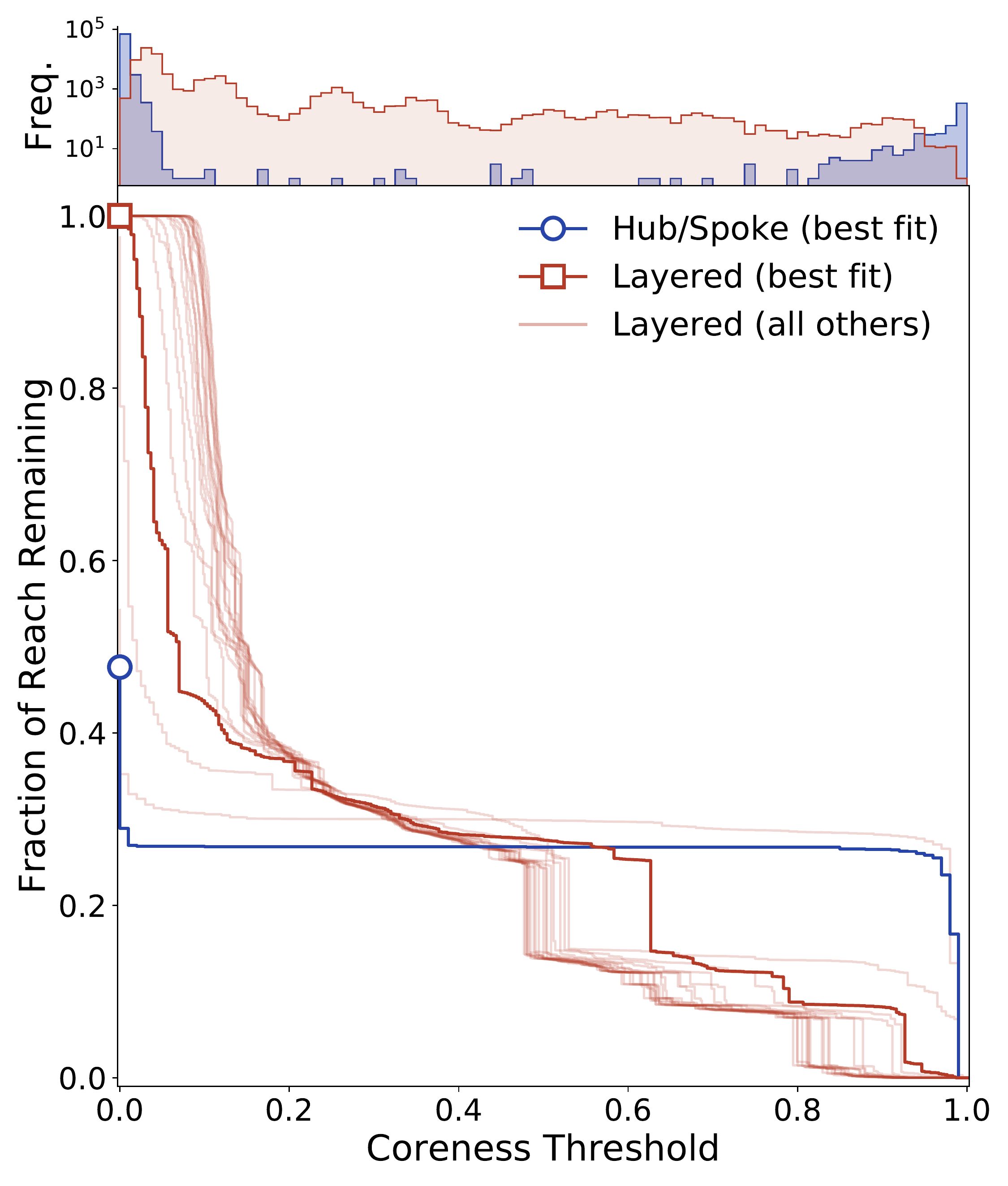}
    \caption{Core-periphery amplification of the hashtag \#MeToo during its first 12 hours of use in October 2017. Reach is measured as the cumulative number of followers among those in the network. Curves show how the fraction of total reach decomposes as the coreness threshold for inclusion into the retweet network is increased. The solid blue curve indicates the best fit hub-and-spoke curve (and best fit overall), the solid red line indicates the best fit layered curve ($\ell=4$ layers), and lighter red lines indicate other layered models with 2 to 20 layers. Markers on the vertical axis indicate the reach after removing nodes with coreness of exactly 0. The histogram above the plot shows the distribution of coreness among nodes in the network for each best fit model.}
    \label{fig-decomposition}
\end{figure}

We iteratively remove individuals from the retweet network according to their coreness, decomposing the network from the periphery to the core, and measure how the hashtag reach varies (see Figure~\ref{fig-decomposition}). We observe that the cumulative reach, the total number of possible followers exposed to the hashtag, declines sharply for both the hub-and-spoke and layered models as the abundance of peripheral amplifiers is removed. However, the reach drops more rapidly for the hub-and-spoke model (MDL = 7.6 bits per edge, the best fit model overall) than any of the layered models and, in particular, the best fit layered model with $\ell=4$ layers (MDL = 9.9 bits per edge). Comparatively, the layered models dramatically underestimate the contribution of the periphery to the early reach of \#MeToo; relative to the hub-and-spoke model, the estimates of reach by the best fit layered model have an error of 63\% at a coreness threshold of 0.1, and an error of 36\% at a threshold of 0.2.

This example illustrates why it is critical to account for the core-periphery typology to make sound network inferences. Qualitatively, the Bayesian block models give us a succinct description of the \#MeToo retweet network, informing us that it is best described as a hub-and-spoke structure that serves to broadcast a small set of core voices, rather than a layered structure with many connections among those disclosing at the periphery.
Quantitatively, using the MDL to select the hub-and-spoke model as the best fit to our network data allows us to confidently estimate the periphery's contribution to the hashtag's reach. This measure can be used to compare across instances of hashtag activism and assess the effectiveness of peripheral amplification, or to develop interventions to counteract amplification manipulation tactics, such as those deployed by social bots and coordinated information operations.

\section{Discussion}

We have presented a typology of core-periphery structure which raises the important distinction between two characterizations: hub-and-spoke and layered. These structures, which are reflected in two of the most widely used core-periphery algorithms \cite{borgatti2000models,seidman1983network}, often yield starkly different descriptions of a network’s core-periphery layout. To elucidate the typology, we have formulated two Bayesian stochastic block models which statistically encode the hub-and-spoke and layered structures. By applying description length as an information-theoretic measure of model fit across a large network database, we have shown empirically that networks express a rich variety of core-periphery structure. Through a case study of online amplification of hashtag activism, we have demonstrated that the choice of core-periphery model used to describe a network affects the substantive interpretation of the network’s structure and function, indicating the need to distinguish between hub-and-spoke and layered structures.

The core-periphery models that we have developed here capture the essential qualitative differences between the hub-and-spoke and layered characterizations. However, they are just a starting point and they offer several paths for extensions and improvements. Our network models do not consider the weights of edges, which could be incorporated to more closely study the cohesiveness of core-periphery structure and how it varies with weight distributions. Our models also do not account for directionality, which can be used to infer distinct in-cores and out-cores \cite{boyd2010computing}. Finally, our models focus on identifying a single core-periphery structure, while a number of networks display multiple core-periphery structure \cite{kojaku2017finding}. Our stochastic block models can be naturally incorporated into more general hierarchical models \cite{peixoto2017nonparametric} for inferring multiple core-periphery pairs, and extended to model directed, weighted edges. The core-periphery models that we have proposed, and the model fit measure for distinguishing them, enable all of these future directions, allowing for more acute studies of core and peripheral dynamics that were not previously possible \cite{csermely2013structure}.

In formulating Bayesian core-periphery block models, we have introduced the layered block model, a new representation of core-periphery structure that is interesting itself. It is amenable to all of the classic network science questions that are usually asked of mesoscale network algorithms \cite{young2018universality}, including questions about its resolution limits \cite{fortunato2007resolution,zhang2015identification} and its interaction with heterogeneous degree distributions \cite{zhang2015identification,karrer2011stochastic}. The layered and hub-and-spoke models do not preclude other formulations of core-periphery structure though, but, rather, point towards the need to formally distinguish between different core-periphery characterizations.

Core-periphery stochastic block models specifically requires us to \emph{restrict} the search space of the statistical inference to smaller subset of possible block matrices. These restriction reduce the expressivity of the model, but, at the same time, allows substantive domain experts to guide models as they apply core-periphery models to network data sets---which is we argue is critical \cite{young2018universality}. For example, we have shown that constrained core-periphery models allow us to better understand dynamics of hashtag activism amplification \cite{barbera2015critical}, a phenomenon that primarily depends on the relationship between the core and the periphery, and not on community structure, disassortativity, or any other mesoscale pattern. Other networked processes that are explained through core-periphery structure, such as brain function and disease spreading, could also benefit from being revisited through the focused lens of our core-periphery typology made possible by our constrained block models. Recognizing that there is no ``ground-truth’’ structure to networks \cite{peel2017ground} and that there are only models that do and do not help us address particular questions, practitioners are uniquely positioned to highlight what questions cannot be answered with the current network models at hand, and where network science can most benefit other disciplines. In tandem, there is a resounding potential for network scientists to fill those methodological gaps, and present models that, themselves, may open doors to new theories and questions. Our core-periphery typology and models clarify the ways in which core-periphery algorithms can be applied to networks, and provide an example of how we, as both domain experts and network scientists, can begin to better align our structural methodology with our substantive questions.

\section{Materials and Methods}
\label{methods}

% ----------------------------------
\subsection{KONECT Network Data}
\label{methods-konect}

The Kolbenz Network Collection (KONECT) \cite{kunegis2013konect} consists of 261 networks and represents a variety of network domains. Networks in the collection may be undirected, directed, or bipartite, and they can contain multiedges and self loops. The edges themselves can be unweighted, weighted, signed, or temporal. We take the following preprocessing steps:
\begin{enumerate}
    \itemsep -0.2em
    \item weighted edges are treated as unweighted,  and all multiedges are collapsed to a single edge;
    \item self-loops are disregarded;
    \item directed edges are treated as undirected;
    \item only the largest weakly connected component is considered.
\end{enumerate}
We exclude all temporal and dynamic networks in KONECT to avoid the ambiguity in choosing a time scale to define static networks. We also exclude all networks that were marked as ``incomplete'' in KONECT \cite{kunegis2013konect}. Finally, we  exclude bipartite networks because they should be modeled with stochastic block models that can account for their special structure and high local density when projected \cite{larremore2014efficiently,gerlach2018network}. After these preprocessing and inclusion criteria, we are left with 142 networks, listed in Appendix~\ref{appendix-konect}. 

We infer block models for all KONECT networks with up to 200,000 nodes, a total of 95 of the 142 networks. The networks with more than 200,000 nodes are concentrated on a small set of network domains: 25\% are social networks, 25\% are hyperlink networks, and 23\% are communication networks. Fitting larger networks is possible, but requires considerable computation.

% ----------------------------------
\subsection{Stochastic Block Model Formulation}
\label{methods-block-models}

Recall that  we let $\A$ be the adjacency matrix of an unweighted, undirected, simple network with $N$ nodes. Nodes are assigned to a fixed number of $B$ blocks, represented by $\theta$, a vector of length $N$ where $\theta_i = r$ indicates that node $i$ belongs to block $r$.  Connections between blocks are specified by the the $B \times B$ matrix $\p$, where $p_{rs}$ is the probability that a node in block $r$ is connected to a node in block $s$. 
The posterior distribution of the parameters $\theta$ and $\p$ given our network data $\A$ is written as
\begin{equation}
    P(\theta, \p \mid \A)
        \propto
            P(\A \mid \theta, \p) P(\theta) P(\p).
\end{equation}
In Section~\ref{results-models} we construct the prior $P(\p)$ on the block connectivity matrix, which is the primary alteration needed for the core-periphery block models:
\begin{align*}
    P_{\mathcal{H}}(\p)  
        &= 3! \cdot  \mathds{1}_{\{0 < p_{22} < p_{12} < p_{11} < 1\}},\\
    P_{\mathcal{L}}(\p) 
        &=   \ell ! \cdot  \mathds{1}_{\{0 < p_{\ell} < p_{\ell-1} < \ldots < p_{1} < 1\}},
\end{align*}
where the leading numerical factors ensure normalization.
Here we think of blocks as layer, so we let $B=\ell$, where $\ell$ is the number of layers.

We otherwise use a standard formulation of the likelihood $P(\A \mid \theta, \p)$ and the block assignments prior $P(\theta)$ \cite{karrer2011stochastic,peixoto2017bayesian}.
The network likelihood rests upon the cornerstone assumption of the stochastic block model: connections are independently generated based only on the block assignments of nodes. Let $m_{rs}$ be the number of edges that exist between blocks $r$ and $s$, and let $M_{rs}$ be the maximum number of edges that could potentially exist between the two blocks. This number equals $n_r n_s$ for two different blocks of size $n_r$ and $n_s$, and to $n_r(n_r-1)/2$ when considering the internal edges of block $r$. The likelihood can be then calculated as the product of independent Bernoulli processes across edges and then aggregated at the block level to yield
\begin{equation}
    P(\A \mid \theta, \p)
        = 
            \prod_{r \leq s}
                 p_{rs}^{m_{rs}} (1 - p_{rs})^{M_{rs} - m_{rs}}.
\end{equation}
The constraints on $\p$ yield a more compact form for this likelihood, see Appendix~\ref{appendix-estimation}.

The last missing piece of the model is the layer assignment prior $P(\theta)$. The prior on $\theta$ can then be expressed in three parts \cite{peixoto2017bayesian}. First, we consider the probability $P(\ell)$ of choosing a particular number of layers $\ell$, which is always $\ell = 2$ for the hub-and-spoke model and a free parameter for the layered model. Next, given the number of layers, we consider the probability $P(\n \mid \ell)$ of drawing a particular sequence of layer sizes $\n=\{n_1,n_2,...n_B\}$. Finally, given the layer sizes, we determine the probability $P(\theta \mid \n)$ of seeing a particular allotment of nodes to layers. All together in notation, the prior on the block assignments $\theta$ is expressed as
\begin{equation}
    P(\theta)
        =
            P(\theta \mid \n) P( \n \mid \ell) P(\ell)
                =
                    \frac{\prod_r n_r !}{N!}
                    \binom{N - 1}{\ell - 1}^{-1}
                    N^{-1}.
\end{equation}

With these three parts of the model specified we can calculate the posterior probability of the model. For more details on the stochastic block model formulation, see refs.~\cite{karrer2011stochastic,peixoto2017bayesian}.
Our estimation methods for the complete resulting models are given Appendix~\ref{appendix-estimation}.

% ----------------------------------
\subsection{Synthetic Networks}
\label{methods-synthetic}

\subsubsection{Discernment Experiment}
\label{methods-3layer}

The first synthetic network experiment tests the ability of the core-periphery models to discern between hub-and-spoke and layered structures (see Fig.~\ref{fig-synthetic-networks}A). We generate networks through the stochastic block model according to block matrices given by
\begin{equation}
    \left[
        \begin{matrix}
            p \gamma & p & p(1 - \delta) + \frac{p}{\gamma} \delta \\
            p & p \delta + \frac{p}{\gamma} (1 - \delta) & \frac{p}{\gamma} \\
            p(1 - \delta) + \frac{p}{\gamma} \delta & \frac{p}{\gamma} & \frac{p}{\gamma}
        \end{matrix}
    \right],
\end{equation}
where $p > 0$ is the baseline density of the network, $\gamma \in [1, 1/p]$ is the structural clarity parameter, and $\delta \in [0, 1]$ is the interpolation parameter. The structural clarity parameter $\gamma$ determines the prevalence of core-periphery structure in the network. When $\gamma = 1$, the network as a whole is simply an Erd\H{os}-R\'enyi random network with density $p$. When $\gamma \gg 1$, the core-periphery structure is well-defined. The interpolation parameter $\delta$ specifies whether a layered or hub-and-spoke core-periphery structure is reflected in the network. When $\delta = 0$, the block densities arrange in such a way that the network is effictively generated from two blocks, and a hub-and-spoke structure is present. When $\delta = 1$, the network exhibits a 3-layer structure. We note that $\delta=1/2$ holds no special meaning in this interpolation: the structure smoothly transitions from one type to the other as $\delta$ is varied.

For the experiment, each synthetic network consists of $n = 10,000$ nodes, divided equally among the three blocks. We set $p = 0.0075$ and generate networks over the parameter ranges $\delta \in [0, 1]$ and $\gamma \in [1, 4]$. See \hyperlink{methods-inference}{Block Model Inference and Parameters} for details on inference of the experimental network structure.

% ----------------------------------------

\subsubsection{Number of Layers Experiment}
\label{methods-merged}

The second synthetic network experiment tests the ability of the layered model to identify the number of layers in synthetic layered networks  (see Fig.~\ref{fig-synthetic-networks}B). We first generate a network $G$ via the layered stochastic block model, where $G$ has $n = 10,000$ nodes evenly split among $\ell = 6$ layers, where layers are connected according to an initial connectivity matrix $\p^{(G)}$. We then consider a new network $G_k$ of the same number of nodes and layers,  but where  $p^{(G_k)}_r = p^{(G)}_r$ for $r < k$, and $p^{(G_k)}_r = q_k$ for $r \geq k$. This is a network where the inner layers have the same density as in $G$, but where the outermost layers are effectively merged because they have the same density $q_k$. We set $q_k$ such that the overall average degree $\kappa$ of $G$ is preserved, i.e. the average degree of $G_k$ is $\kappa$ for all $k$. The merged layers density $q_k$ preserving $\kappa$ is given by
\begin{equation}
q_k
    =
        \frac{
            \binom{n}{2} \sum_{r=k}^\ell p^{(G)}_r
                +n^2 \sum_{r = k}^\ell (r - 1) p^{(G)}_r
        }
        {
            \binom{n}{2} (\ell - k + 1)
                + n^2 \sum_{r=k}^\ell (r -1 )
        }.
\end{equation}

For each choice of $k \in [2,6]$, we make 10 networks generated from the same block matrix. We define the block matrix $\p^{(G)}$ of the original network such that $p_1 = 0.002$, $p_6 = 0.1$, and $p_r$ for $1 < r < 6$ is geometrically distributed between $p_1$ and $p_6$. See \hyperlink{methods-inference}{Block Model Inference and Parameters} for details on inference of the experimental network structure.

% ----------------------------------
\subsection{Hashtag Activism Case Study}
\label{methods-metoo}

For the hashtag activism case study, we consider all of the tweets containing the hashtag \#MeToo that were posted 12 hours after actress Alyssa Milano's ``me too'' tweet, which catalyzed the hashtag movement on October 15th, 2017 (see ref.~\cite{gallagher2019reclaiming} for further data details). In that time frame, there were 208,926 tweets. We construct a retweet network from these tweets by representing individuals as nodes and retweets as edges. For the purpose of core-periphery modeling, we treat edges as undirected and unweighted and remove self-loops from the network. We model the largest weakly connected component of the network \cite{newman2018networks}, which consists of 74,214 nodes and 130,277 edges.

We measure the ``coreness'' of each individual by taking into account all the potential core-periphery descriptions identified by a model. We can consider the average block or layer $\theta_i$ of a node as a measure of its distance to the core of the network, and use that to define coreness as
\begin{equation}
c_i
    = 
        1 - 
        \frac{1}{\ell}
        \sum_{r = 1}^\ell
            r \, P(\theta_i = r \mid \A).
\end{equation}
In this expression, $\ell$ is the number of blocks, and $P(\theta_i = r \mid \A)$ 
is the probability that node $i$ takes on block assignment $r$. The latter probability is the marginal distribution of $\theta_i$, formally defined as
\begin{equation}
P(\theta_i = r \mid \A)
    =
        \sum_{\theta}
            P(\theta \mid A) \, \mathds{1}_{\bigl\{ \theta_i^{(t)} = r \bigr\}}.
\end{equation}
Coreness varies between 0 and 1, where an individual positioned consistently in the core will have a higher coreness score.

% ----------------------------------
\subsection{Block Model Inference and Parameters}
\label{methods-inference}

For the discernment experiment, we run the hub-and-spoke and layered models 3 times each for each $(\gamma, \delta)$ parameter tuple. For each model, we use the best model according to the MDL. For each run of each model, we sweep over 250 Gibbs samples and let each MCMC simulation run for 10 times the number of nodes in the network (see Appendix~\ref{appendix-estimation} for numerical details). We use samples from the second half of the Gibbs sampling chain to infer the parameters $\hat{\theta}$, the most probable block labels. We use $10^7$ samples to approximate the MDL  (see Appendix~\ref{appendix-model-fit} for numerical details).

For the layered experiment, we consider $L$, the actual number of layers in each synthetic network, and $\ell$, the fixed parameter in the layered model. For each $L$, there are $N_L = 10$ networks. For each of those networks, we run the layered models 3 times and choose the best model from those 3 runs according to the MDL. We then average the MDL over the $N_L$ networks to get the average MDL per $(L, \ell)$ pair. We perform inference similar to the discernment experiment, but instead use 20 steps per node for the MCMC chains. To account for more layers than the previous experiment, we use $10^8$ samples to approximate the MDL.

For each KONECT network and the \#MeToo case study network, we run both the hub-and-spoke and layered models 3 times each. For each model, we choose the best run, as determined by the MDL. For the KONECT networks, we run layered models for $\ell \in [2, 10]$ and use the model with the best MDL across all choices of $\ell$. For the \#MeToo network, we vary $\ell$ in the range $\ell \in [2, 20]$ and choose the best model overall (thick red line in Fig.~\ref{fig-decomposition}) and for each individual choice of $\ell$ (light red lines in Fig~\ref{fig-decomposition}) according to the MDL. We use 200 Gibbs samples for the KONECT and \#MeToo models, and infer partitions according to the second half of each chain. For the MCMC chains, we use 10 steps per node.  We use $10^8$ samples to estimate the MDL.

The Python code  for inferring the hub-and-spoke and layered core-periphery models and evaluating their model fit is freely available online at \url{https://github.com/ryanjgallagher/core_periphery_sbm}.

\vspace{1em}

\section*{Acknowledgements}

We thank Daniel Larremore for the initial inspiration and depiction of the layered block model presented in this work. We also thank Nicholas Beauchamp, Aaron Clauset, Leo Torres, and Jessica Davis for their conversations early in the project, and Brennan Klein for his assistance and advice on the data visualization. This work was  supported in part by equipment and computing resources from NVIDIA Corporation and Northeastern University's Discovery computing cluster. J-GY was supported by a James S. McDonnell Foundation Postdoctoral Fellowship Award.

The KONECT data set used in this work is \href{http://konect.uni-koblenz.de/}{freely available online}. The Twitter data underlying the \#MeToo case study is available at the University of Michigan's \href{https://www.icpsr.umich.edu/web/ICPSR/studies/37447}{Inter-University Consortium for Political and Social Research} upon submission and acceptance of a Restricted Data Use Agreement.

% ---------------------------------------------------------
% --------------------- Bibliography ----------------------
% ---------------------------------------------------------

\bibliographystyle{unsrt}
\bibliography{bib}

% ---------------------------------------------------------
% ----------------------- Appendix ------------------------
% ---------------------------------------------------------

\clearpage
\onecolumngrid

\appendix

\renewcommand\thefigure{\thesection\arabic{figure}}
\renewcommand\thetable{\thesection\arabic{table}}   

\setcounter{figure}{0}
\section{KONECT Core-Periphery Modeling}
\label{appendix-konect}

\subsection{Core-Periphery Statistics}

\begingroup
\renewcommand\arraystretch{1.0}
\setlength{\tabcolsep}{9pt}
\begin{longtable}{llllllll}
\textbf{Dataset} & \textbf{Domain} & $\mathbf{N}$ & $\mathbf{\langle k \rangle}$ & \textbf{VI} & $\mathbf{\Sigma_\mathcal{H} / m}$  & $\mathbf{\Sigma_\mathcal{L} / m}$ & $\mathbf{L}$ \\ \hline
\scriptsize Rhesus & \scriptsize Animal & \scriptsize 16 & \scriptsize 8.60 & \scriptsize 2.24 & \scriptsize \textbf{0.55} & \scriptsize 0.56 & \scriptsize 2 \\ \hline
\scriptsize Kangaroo & \scriptsize Animal & \scriptsize 17 & \scriptsize 10.70 & \scriptsize 1.40 & \scriptsize 0.37 & \scriptsize \textbf{0.37} & \scriptsize 2 \\ \hline
\scriptsize Zebra & \scriptsize Animal & \scriptsize 23 & \scriptsize 9.10 & \scriptsize 1.85 & \scriptsize 0.48 & \scriptsize \textbf{0.47} & \scriptsize 2 \\ \hline
\scriptsize Bison & \scriptsize Animal & \scriptsize 26 & \scriptsize 17.10 & \scriptsize 2.00 & \scriptsize 0.36 & \scriptsize \textbf{0.36} & \scriptsize 2 \\ \hline
\scriptsize Cattle & \scriptsize Animal & \scriptsize 28 & \scriptsize 14.60 & \scriptsize 1.82 & \scriptsize 0.54 & \scriptsize \textbf{0.54} & \scriptsize 2 \\ \hline
\scriptsize Sheep & \scriptsize Animal & \scriptsize 28 & \scriptsize 16.80 & \scriptsize 1.58 & \scriptsize 0.43 & \scriptsize \textbf{0.43} & \scriptsize 2 \\ \hline
\scriptsize Hens & \scriptsize Animal & \scriptsize 32 & \scriptsize 31.00 & \scriptsize 0.20 & \scriptsize 0.02 & \scriptsize \textbf{0.02} & \scriptsize 2 \\ \hline
\scriptsize Dolphins & \scriptsize Animal & \scriptsize 62 & \scriptsize 5.10 & \scriptsize 2.16 & \scriptsize 1.52 & \scriptsize \textbf{1.51} & \scriptsize 2 \\ \hline
\scriptsize Macaques & \scriptsize Animal & \scriptsize 62 & \scriptsize 37.60 & \scriptsize 1.57 & \scriptsize 0.46 & \scriptsize \textbf{0.46} & \scriptsize 2 \\ \hline
\scriptsize arXiv astro-ph & \scriptsize Authorship & \scriptsize 17,903 & \scriptsize 22.00 & \scriptsize 5.05 & \scriptsize 2.95 & \scriptsize \textbf{2.81} & \scriptsize 9 \\ \hline
\scriptsize arXiv hep-th & \scriptsize Authorship & \scriptsize 22,721 & \scriptsize 215.20 & \scriptsize 7.85 & \scriptsize 1.93 & \scriptsize \textbf{1.73} & \scriptsize 10 \\ \hline
\scriptsize arXiv hep-ph & \scriptsize Authorship & \scriptsize 28,045 & \scriptsize 224.50 & \scriptsize 7.95 & \scriptsize 2.13 & \scriptsize \textbf{1.97} & \scriptsize 10 \\ \hline
\scriptsize Dblp Co-Authorship & \scriptsize Authorship & \scriptsize 317,080 & \scriptsize 6.60 & \scriptsize 3.23 & \scriptsize  & \scriptsize  & \scriptsize  \\ \hline
\scriptsize Dblp & \scriptsize Citation & \scriptsize 12,495 & \scriptsize 7.90 & \scriptsize 3.01 & \scriptsize 3.17 & \scriptsize \textbf{3.17} & \scriptsize 5 \\ \hline
\scriptsize Cora Citation & \scriptsize Citation & \scriptsize 23,166 & \scriptsize 7.70 & \scriptsize 3.22 & \scriptsize 3.69 & \scriptsize \textbf{3.66} & \scriptsize 4 \\ \hline
\scriptsize arXiv hep-th & \scriptsize Citation & \scriptsize 27,400 & \scriptsize 25.70 & \scriptsize 4.95 & \scriptsize 3.11 & \scriptsize \textbf{3.00} & \scriptsize 8 \\ \hline
\scriptsize arXiv hep-ph & \scriptsize Citation & \scriptsize 34,401 & \scriptsize 24.50 & \scriptsize 4.80 & \scriptsize 3.33 & \scriptsize \textbf{3.23} & \scriptsize 7 \\ \hline
\scriptsize Citeseer & \scriptsize Citation & \scriptsize 365,154 & \scriptsize 9.40 & \scriptsize 3.53 & \scriptsize  & \scriptsize  & \scriptsize  \\ \hline
\scriptsize US Patents & \scriptsize Citation & \scriptsize 3,764,117 & \scriptsize 8.80 & \scriptsize 3.39 & \scriptsize  & \scriptsize  & \scriptsize  \\ \hline
\scriptsize Manufacturing Emails & \scriptsize Communication & \scriptsize 167 & \scriptsize 38.90 & \scriptsize 3.48 & \scriptsize 0.75 & \scriptsize \textbf{0.70} & \scriptsize 6 \\ \hline
\scriptsize Dnc Emails Co-Recipients & \scriptsize Communication & \scriptsize 849 & \scriptsize 24.50 & \scriptsize 3.91 & \scriptsize 1.21 & \scriptsize \textbf{1.09} & \scriptsize 7 \\ \hline
\scriptsize U. Rovira I Virgili & \scriptsize Communication & \scriptsize 1,133 & \scriptsize 9.60 & \scriptsize 3.49 & \scriptsize 2.32 & \scriptsize \textbf{2.30} & \scriptsize 4 \\ \hline
\scriptsize Hamsterster Friendships & \scriptsize Communication & \scriptsize 1,788 & \scriptsize 14.00 & \scriptsize 4.10 & \scriptsize 2.17 & \scriptsize \textbf{2.14} & \scriptsize 7 \\ \hline
\scriptsize Dnc Emails & \scriptsize Communication & \scriptsize 1,833 & \scriptsize 4.80 & \scriptsize 2.12 & \scriptsize \textbf{2.05} & \scriptsize 2.38 & \scriptsize 5 \\ \hline
\scriptsize Uc Irvine Messages & \scriptsize Communication & \scriptsize 1,893 & \scriptsize 14.60 & \scriptsize 4.07 & \scriptsize 2.10 & \scriptsize \textbf{2.09} & \scriptsize 6 \\ \hline
\scriptsize Hamsterster Full & \scriptsize Communication & \scriptsize 2,000 & \scriptsize 16.10 & \scriptsize 4.47 & \scriptsize 2.23 & \scriptsize \textbf{2.14} & \scriptsize 7 \\ \hline
\scriptsize Facebook (Nips) & \scriptsize Communication & \scriptsize 2,888 & \scriptsize 2.10 & \scriptsize 0.25 & \scriptsize \textbf{1.41} & \scriptsize 3.56 & \scriptsize 2 \\ \hline
\scriptsize Advogato & \scriptsize Communication & \scriptsize 5,042 & \scriptsize 15.60 & \scriptsize 4.14 & \scriptsize 2.43 & \scriptsize \textbf{2.38} & \scriptsize 7 \\ \hline
\scriptsize Pretty Good Privacy & \scriptsize Communication & \scriptsize 10,680 & \scriptsize 4.60 & \scriptsize 2.44 & \scriptsize 3.38 & \scriptsize \textbf{3.27} & \scriptsize 6 \\ \hline
\scriptsize Twitter Lists & \scriptsize Communication & \scriptsize 22,322 & \scriptsize 2.90 & \scriptsize 1.23 & \scriptsize \textbf{3.23} & \scriptsize 4.07 & \scriptsize 3 \\ \hline
\scriptsize Google+ & \scriptsize Communication & \scriptsize 23,613 & \scriptsize 3.30 & \scriptsize 1.57 & \scriptsize \textbf{2.29} & \scriptsize 3.96 & \scriptsize 3 \\ \hline
\scriptsize Linux Kernel Mailing List Replies & \scriptsize Communication & \scriptsize 24,567 & \scriptsize 12.90 & \scriptsize 3.36 & \scriptsize 2.54 & \scriptsize \textbf{2.46} & \scriptsize 8 \\ \hline
\scriptsize Digg & \scriptsize Communication & \scriptsize 29,652 & \scriptsize 5.70 & \scriptsize 2.65 & \scriptsize \textbf{3.73} & \scriptsize 3.79 & \scriptsize 4 \\ \hline
\scriptsize Facebook Wall Posts & \scriptsize Communication & \scriptsize 43,953 & \scriptsize 8.30 & \scriptsize 3.37 & \scriptsize 3.82 & \scriptsize \textbf{3.74} & \scriptsize 6 \\ \hline
\scriptsize Slashdot Threads & \scriptsize Communication & \scriptsize 51,083 & \scriptsize 4.60 & \scriptsize 2.08 & \scriptsize \textbf{3.77} & \scriptsize 3.94 & \scriptsize 5 \\ \hline
\scriptsize Brightkite & \scriptsize Communication & \scriptsize 56,739 & \scriptsize 7.50 & \scriptsize 3.08 & \scriptsize 3.73 & \scriptsize \textbf{3.62} & \scriptsize 7 \\ \hline
\scriptsize Epinions & \scriptsize Communication & \scriptsize 75,877 & \scriptsize 10.70 & \scriptsize 2.97 & \scriptsize 3.19 & \scriptsize \textbf{3.07} & \scriptsize 8 \\ \hline
\scriptsize Slashdot Zoo & \scriptsize Communication & \scriptsize 79,116 & \scriptsize 11.80 & \scriptsize 3.52 & \scriptsize 3.40 & \scriptsize \textbf{3.35} & \scriptsize 8 \\ \hline
\scriptsize Enron & \scriptsize Communication & \scriptsize 84,384 & \scriptsize 7.00 & \scriptsize 2.24 & \scriptsize \textbf{3.30} & \scriptsize 3.38 & \scriptsize 7 \\ \hline
\scriptsize Wikipedia Threads (De) & \scriptsize Communication & \scriptsize 89,146 & \scriptsize 16.30 & \scriptsize 3.63 & \scriptsize 2.93 & \scriptsize \textbf{2.84} & \scriptsize 9 \\ \hline
\scriptsize Livemocha & \scriptsize Communication & \scriptsize 104,103 & \scriptsize 42.10 & \scriptsize 5.74 & \scriptsize 3.16 & \scriptsize \textbf{3.11} & \scriptsize 10 \\ \hline
\scriptsize Wikipedia Conflict & \scriptsize Communication & \scriptsize 113,123 & \scriptsize 35.80 & \scriptsize 4.95 & \scriptsize 3.04 & \scriptsize \textbf{2.87} & \scriptsize 10 \\ \hline
\scriptsize Epinions Trust & \scriptsize Communication & \scriptsize 119,130 & \scriptsize 11.80 & \scriptsize 3.05 & \scriptsize 3.22 & \scriptsize \textbf{3.11} & \scriptsize 9 \\ \hline
\scriptsize Wikisigned & \scriptsize Communication & \scriptsize 137,740 & \scriptsize 10.40 & \scriptsize 3.15 & \scriptsize 3.58 & \scriptsize \textbf{3.54} & \scriptsize 9 \\ \hline
\scriptsize Catster Friendships & \scriptsize Communication & \scriptsize 148,826 & \scriptsize 73.20 & \scriptsize 6.37 & \scriptsize \textbf{2.57} & \scriptsize 2.76 & \scriptsize 10 \\ \hline
\scriptsize Douban & \scriptsize Communication & \scriptsize 154,908 & \scriptsize 4.20 & \scriptsize 1.85 & \scriptsize \textbf{3.96} & \scriptsize 4.43 & \scriptsize 5 \\ \hline
\scriptsize Gowalla & \scriptsize Communication & \scriptsize 196,591 & \scriptsize 9.70 & \scriptsize 3.54 & \scriptsize 4.15 & \scriptsize \textbf{4.01} & \scriptsize 8 \\ \hline
\scriptsize Libimseti.cz & \scriptsize Communication & \scriptsize 220,970 & \scriptsize 156.00 & \scriptsize 7.41 & \scriptsize  & \scriptsize  & \scriptsize  \\ \hline
\scriptsize Wikipedia Talk (Dutch) & \scriptsize Communication & \scriptsize 224,185 & \scriptsize 4.60 & \scriptsize 1.98 & \scriptsize  & \scriptsize  & \scriptsize  \\ \hline
\scriptsize Dogster Friendships & \scriptsize Communication & \scriptsize 426,485 & \scriptsize 40.10 & \scriptsize 5.64 & \scriptsize  & \scriptsize  & \scriptsize  \\ \hline
\scriptsize Wikipedia Talk (Russian) & \scriptsize Communication & \scriptsize 449,042 & \scriptsize 3.80 & \scriptsize 1.59 & \scriptsize  & \scriptsize  & \scriptsize  \\ \hline
\scriptsize Twitter (ICWSM) & \scriptsize Communication & \scriptsize 465,017 & \scriptsize 3.60 & \scriptsize 1.44 & \scriptsize  & \scriptsize  & \scriptsize  \\ \hline
\scriptsize Wikipedia Talk (Spanish) & \scriptsize Communication & \scriptsize 476,465 & \scriptsize 4.50 & \scriptsize 1.82 & \scriptsize  & \scriptsize  & \scriptsize  \\ \hline
\scriptsize Wikipedia Talk (German) & \scriptsize Communication & \scriptsize 505,468 & \scriptsize 6.00 & \scriptsize 2.17 & \scriptsize  & \scriptsize  & \scriptsize  \\ \hline
\scriptsize Wikipedia Talk (Portuguese) & \scriptsize Communication & \scriptsize 534,618 & \scriptsize 5.20 & \scriptsize 2.13 & \scriptsize  & \scriptsize  & \scriptsize  \\ \hline
\scriptsize Catster/Dogster Family/Friendships & \scriptsize Communication & \scriptsize 601,213 & \scriptsize 52.10 & \scriptsize 5.95 & \scriptsize  & \scriptsize  & \scriptsize  \\ \hline
\scriptsize Wikipedia Talk (Italian) & \scriptsize Communication & \scriptsize 862,214 & \scriptsize 3.50 & \scriptsize 1.44 & \scriptsize  & \scriptsize  & \scriptsize  \\ \hline
\scriptsize Wikipedia Talk (Arabic) & \scriptsize Communication & \scriptsize 1,095,524 & \scriptsize 2.80 & \scriptsize 1.17 & \scriptsize  & \scriptsize  & \scriptsize  \\ \hline
\scriptsize Wikipedia Talk (Chinese) & \scriptsize Communication & \scriptsize 1,217,365 & \scriptsize 2.80 & \scriptsize 1.06 & \scriptsize  & \scriptsize  & \scriptsize  \\ \hline
\scriptsize Hyves & \scriptsize Communication & \scriptsize 1,402,673 & \scriptsize 4.00 & \scriptsize 1.98 & \scriptsize  & \scriptsize  & \scriptsize  \\ \hline
\scriptsize Wikipedia Talk (French) & \scriptsize Communication & \scriptsize 1,409,540 & \scriptsize 3.20 & \scriptsize 1.28 & \scriptsize  & \scriptsize  & \scriptsize  \\ \hline
\scriptsize Flickr Links & \scriptsize Communication & \scriptsize 1,624,991 & \scriptsize 19.00 & \scriptsize 3.05 & \scriptsize  & \scriptsize  & \scriptsize  \\ \hline
\scriptsize Pokec & \scriptsize Communication & \scriptsize 1,632,803 & \scriptsize 27.30 & \scriptsize 4.93 & \scriptsize  & \scriptsize  & \scriptsize  \\ \hline
\scriptsize Flickr & \scriptsize Communication & \scriptsize 2,173,370 & \scriptsize 20.90 & \scriptsize 3.12 & \scriptsize  & \scriptsize  & \scriptsize  \\ \hline
\scriptsize Wikipedia (English) & \scriptsize Communication & \scriptsize 2,388,953 & \scriptsize 3.90 & \scriptsize 1.47 & \scriptsize  & \scriptsize  & \scriptsize  \\ \hline
\scriptsize Flixster & \scriptsize Communication & \scriptsize 2,523,386 & \scriptsize 6.30 & \scriptsize 2.35 & \scriptsize  & \scriptsize  & \scriptsize  \\ \hline
\scriptsize Wikipedia Talk (English) & \scriptsize Communication & \scriptsize 2,859,574 & \scriptsize 5.70 & \scriptsize 2.18 & \scriptsize  & \scriptsize  & \scriptsize  \\ \hline
\scriptsize Youtube & \scriptsize Communication & \scriptsize 3,216,075 & \scriptsize 5.80 & \scriptsize 2.36 & \scriptsize  & \scriptsize  & \scriptsize  \\ \hline
\scriptsize Livejournal & \scriptsize Communication & \scriptsize 4,843,953 & \scriptsize 17.70 & \scriptsize 4.39 & \scriptsize  & \scriptsize  & \scriptsize  \\ \hline
\scriptsize Livejournal Links & \scriptsize Communication & \scriptsize 5,189,808 & \scriptsize 18.80 & \scriptsize 4.51 & \scriptsize  & \scriptsize  & \scriptsize  \\ \hline
\scriptsize Route Views & \scriptsize Computer & \scriptsize 6,474 & \scriptsize 3.90 & \scriptsize 1.91 & \scriptsize \textbf{2.82} & \scriptsize 3.45 & \scriptsize 4 \\ \hline
\scriptsize Caida & \scriptsize Computer & \scriptsize 26,475 & \scriptsize 4.00 & \scriptsize 1.93 & \scriptsize \textbf{3.27} & \scriptsize 3.95 & \scriptsize 4 \\ \hline
\scriptsize Internet Topology & \scriptsize Computer & \scriptsize 34,761 & \scriptsize 6.20 & \scriptsize 2.36 & \scriptsize \textbf{3.01} & \scriptsize 3.29 & \scriptsize 6 \\ \hline
\scriptsize Gnutella & \scriptsize Computer & \scriptsize 62,561 & \scriptsize 4.70 & \scriptsize 2.04 & \scriptsize \textbf{4.21} & \scriptsize 4.34 & \scriptsize 3 \\ \hline
\scriptsize Skitter & \scriptsize Computer & \scriptsize 1,694,616 & \scriptsize 13.10 & \scriptsize 4.02 & \scriptsize  & \scriptsize  & \scriptsize  \\ \hline
\scriptsize Blogs & \scriptsize Hyperlink & \scriptsize 1,222 & \scriptsize 27.40 & \scriptsize 4.99 & \scriptsize 1.66 & \scriptsize \textbf{1.59} & \scriptsize 9 \\ \hline
\scriptsize Foldoc & \scriptsize Hyperlink & \scriptsize 13,356 & \scriptsize 13.70 & \scriptsize 2.33 & \scriptsize \textbf{3.32} & \scriptsize 3.33 & \scriptsize 4 \\ \hline
\scriptsize Google.com Internal & \scriptsize Hyperlink & \scriptsize 15,763 & \scriptsize 18.90 & \scriptsize 4.07 & \scriptsize \textbf{2.74} & \scriptsize 2.75 & \scriptsize 10 \\ \hline
\scriptsize Stanford & \scriptsize Hyperlink & \scriptsize 255,265 & \scriptsize 15.20 & \scriptsize 4.20 & \scriptsize  & \scriptsize  & \scriptsize  \\ \hline
\scriptsize Notre Dame & \scriptsize Hyperlink & \scriptsize 325,729 & \scriptsize 6.70 & \scriptsize 2.72 & \scriptsize  & \scriptsize  & \scriptsize  \\ \hline
\scriptsize Baidu Related Pages & \scriptsize Hyperlink & \scriptsize 372,840 & \scriptsize 12.40 & \scriptsize 3.18 & \scriptsize  & \scriptsize  & \scriptsize  \\ \hline
\scriptsize Berkeley/Stanford & \scriptsize Hyperlink & \scriptsize 654,782 & \scriptsize 20.10 & \scriptsize 4.67 & \scriptsize  & \scriptsize  & \scriptsize  \\ \hline
\scriptsize Google & \scriptsize Hyperlink & \scriptsize 855,802 & \scriptsize 10.00 & \scriptsize 3.87 & \scriptsize  & \scriptsize  & \scriptsize  \\ \hline
\scriptsize Trec Wt10G & \scriptsize Hyperlink & \scriptsize 1,458,316 & \scriptsize 8.50 & \scriptsize 3.52 & \scriptsize  & \scriptsize  & \scriptsize  \\ \hline
\scriptsize Wikipedia Links (Polish) & \scriptsize Hyperlink & \scriptsize 1,529,116 & \scriptsize 55.20 & \scriptsize 6.16 & \scriptsize  & \scriptsize  & \scriptsize  \\ \hline
\scriptsize Wikipedia Links (Portuguese) & \scriptsize Hyperlink & \scriptsize 1,602,960 & \scriptsize 48.20 & \scriptsize 5.58 & \scriptsize  & \scriptsize  & \scriptsize  \\ \hline
\scriptsize Wikipedia (English) & \scriptsize Hyperlink & \scriptsize 1,870,521 & \scriptsize 39.10 & \scriptsize 5.69 & \scriptsize  & \scriptsize  & \scriptsize  \\ \hline
\scriptsize Hudong Internal Links & \scriptsize Hyperlink & \scriptsize 1,962,418 & \scriptsize 14.70 & \scriptsize 4.15 & \scriptsize  & \scriptsize  & \scriptsize  \\ \hline
\scriptsize Baidu Internal Links & \scriptsize Hyperlink & \scriptsize 2,107,689 & \scriptsize 16.10 & \scriptsize 4.12 & \scriptsize  & \scriptsize  & \scriptsize  \\ \hline
\scriptsize Hudong Related Pages & \scriptsize Hyperlink & \scriptsize 2,415,542 & \scriptsize 15.50 & \scriptsize 1.94 & \scriptsize  & \scriptsize  & \scriptsize  \\ \hline
\scriptsize Contiguous Usa & \scriptsize Infrastructure & \scriptsize 49 & \scriptsize 4.40 & \scriptsize 1.52 & \scriptsize 1.53 & \scriptsize \textbf{1.53} & \scriptsize 2 \\ \hline
\scriptsize Chicago & \scriptsize Infrastructure & \scriptsize 823 & \scriptsize 2.00 & \scriptsize 0.08 & \scriptsize \textbf{2.56} & \scriptsize 3.15 & \scriptsize 2 \\ \hline
\scriptsize Euroroad & \scriptsize Infrastructure & \scriptsize 1,039 & \scriptsize 2.50 & \scriptsize 0.94 & \scriptsize \textbf{3.28} & \scriptsize 3.29 & \scriptsize 2 \\ \hline
\scriptsize Air Traffic Control & \scriptsize Infrastructure & \scriptsize 1,226 & \scriptsize 3.90 & \scriptsize 1.73 & \scriptsize \textbf{2.85} & \scriptsize 2.87 & \scriptsize 2 \\ \hline
\scriptsize US Airports & \scriptsize Infrastructure & \scriptsize 1,572 & \scriptsize 21.90 & \scriptsize 4.55 & \scriptsize 1.50 & \scriptsize \textbf{1.36} & \scriptsize 10 \\ \hline
\scriptsize Openflights & \scriptsize Infrastructure & \scriptsize 2,905 & \scriptsize 10.80 & \scriptsize 3.63 & \scriptsize 2.22 & \scriptsize \textbf{2.14} & \scriptsize 6 \\ \hline
\scriptsize Openflights & \scriptsize Infrastructure & \scriptsize 3,397 & \scriptsize 11.30 & \scriptsize 3.69 & \scriptsize 2.22 & \scriptsize \textbf{2.15} & \scriptsize 7 \\ \hline
\scriptsize US Power Grid & \scriptsize Infrastructure & \scriptsize 4,941 & \scriptsize 2.70 & \scriptsize 1.21 & \scriptsize 3.70 & \scriptsize \textbf{3.70} & \scriptsize 2 \\ \hline
\scriptsize Pennsylvania & \scriptsize Infrastructure & \scriptsize 1,087,562 & \scriptsize 2.80 & \scriptsize 0.73 & \scriptsize  & \scriptsize  & \scriptsize  \\ \hline
\scriptsize Texas & \scriptsize Infrastructure & \scriptsize 1,351,137 & \scriptsize 2.80 & \scriptsize 0.75 & \scriptsize  & \scriptsize  & \scriptsize  \\ \hline
\scriptsize California & \scriptsize Infrastructure & \scriptsize 1,957,027 & \scriptsize 2.80 & \scriptsize 0.72 & \scriptsize  & \scriptsize  & \scriptsize  \\ \hline
\scriptsize Windsurfers & \scriptsize Interaction & \scriptsize 43 & \scriptsize 15.60 & \scriptsize 2.72 & \scriptsize \textbf{0.74} & \scriptsize 0.75 & \scriptsize 2 \\ \hline
\scriptsize Train Bombing & \scriptsize Interaction & \scriptsize 64 & \scriptsize 7.60 & \scriptsize 3.06 & \scriptsize 1.14 & \scriptsize \textbf{1.13} & \scriptsize 2 \\ \hline
\scriptsize Reality Mining & \scriptsize Interaction & \scriptsize 96 & \scriptsize 52.90 & \scriptsize 3.08 & \scriptsize 0.44 & \scriptsize \textbf{0.41} & \scriptsize 4 \\ \hline
\scriptsize Hypertext 2009 & \scriptsize Interaction & \scriptsize 113 & \scriptsize 38.90 & \scriptsize 3.24 & \scriptsize 0.73 & \scriptsize \textbf{0.73} & \scriptsize 4 \\ \hline
\scriptsize Haggle & \scriptsize Interaction & \scriptsize 274 & \scriptsize 15.50 & \scriptsize 2.97 & \scriptsize \textbf{0.77} & \scriptsize 0.80 & \scriptsize 5 \\ \hline
\scriptsize Infectious & \scriptsize Interaction & \scriptsize 410 & \scriptsize 13.50 & \scriptsize 3.83 & \scriptsize 1.91 & \scriptsize \textbf{1.73} & \scriptsize 5 \\ \hline
\scriptsize Chess & \scriptsize Interaction & \scriptsize 7,115 & \scriptsize 15.70 & \scriptsize 4.37 & \scriptsize 2.69 & \scriptsize \textbf{2.57} & \scriptsize 9 \\ \hline
\scriptsize Prosper Loans & \scriptsize Interaction & \scriptsize 89,171 & \scriptsize 74.70 & \scriptsize 6.48 & \scriptsize 3.08 & \scriptsize \textbf{2.95} & \scriptsize 10 \\ \hline
\scriptsize David Copperfield & \scriptsize Lexical & \scriptsize 112 & \scriptsize 7.60 & \scriptsize 2.81 & \scriptsize \textbf{1.49} & \scriptsize 1.53 & \scriptsize 2 \\ \hline
\scriptsize Bible & \scriptsize Lexical & \scriptsize 1,707 & \scriptsize 10.60 & \scriptsize 3.44 & \scriptsize \textbf{2.31} & \scriptsize 2.32 & \scriptsize 5 \\ \hline
\scriptsize Edinburgh Associative Thesaurus & \scriptsize Lexical & \scriptsize 23,132 & \scriptsize 25.70 & \scriptsize 4.05 & \scriptsize 2.84 & \scriptsize \textbf{2.80} & \scriptsize 7 \\ \hline
\scriptsize Wordnet & \scriptsize Lexical & \scriptsize 145,145 & \scriptsize 9.00 & \scriptsize 3.59 & \scriptsize \textbf{4.29} & \scriptsize 4.30 & \scriptsize 6 \\ \hline
\scriptsize Yahoo Advertisers & \scriptsize Lexical & \scriptsize 653,260 & \scriptsize 9.00 & \scriptsize 2.55 & \scriptsize  & \scriptsize  & \scriptsize  \\ \hline
\scriptsize Pdzbase & \scriptsize Metabolic & \scriptsize 161 & \scriptsize 2.60 & \scriptsize 1.58 & \scriptsize \textbf{2.07} & \scriptsize 2.23 & \scriptsize 2 \\ \hline
\scriptsize Caenorhabditis Elegans & \scriptsize Metabolic & \scriptsize 453 & \scriptsize 8.90 & \scriptsize 3.07 & \scriptsize \textbf{1.72} & \scriptsize 1.96 & \scriptsize 4 \\ \hline
\scriptsize Protein & \scriptsize Metabolic & \scriptsize 1,458 & \scriptsize 2.70 & \scriptsize 1.43 & \scriptsize \textbf{3.03} & \scriptsize 3.15 & \scriptsize 2 \\ \hline
\scriptsize Human Protein (Stelzl) & \scriptsize Metabolic & \scriptsize 1,615 & \scriptsize 3.80 & \scriptsize 2.09 & \scriptsize \textbf{2.67} & \scriptsize 2.81 & \scriptsize 3 \\ \hline
\scriptsize Human Protein (Vidal) & \scriptsize Metabolic & \scriptsize 2,783 & \scriptsize 4.30 & \scriptsize 2.31 & \scriptsize \textbf{2.97} & \scriptsize 3.07 & \scriptsize 3 \\ \hline
\scriptsize Reactome & \scriptsize Metabolic & \scriptsize 5,973 & \scriptsize 48.80 & \scriptsize 5.42 & \scriptsize 2.04 & \scriptsize \textbf{1.68} & \scriptsize 10 \\ \hline
\scriptsize Les Mis\'erables & \scriptsize Miscellaneous & \scriptsize 77 & \scriptsize 6.60 & \scriptsize 3.20 & \scriptsize 1.31 & \scriptsize \textbf{1.31} & \scriptsize 4 \\ \hline
\scriptsize Flickr & \scriptsize Miscellaneous & \scriptsize 105,722 & \scriptsize 43.80 & \scriptsize 2.41 & \scriptsize 2.69 & \scriptsize \textbf{2.43} & \scriptsize 8 \\ \hline
\scriptsize Amazon (Mds) & \scriptsize Miscellaneous & \scriptsize 334,863 & \scriptsize 5.50 & \scriptsize 2.20 & \scriptsize  & \scriptsize  & \scriptsize  \\ \hline
\scriptsize Actor Collaborations & \scriptsize Miscellaneous & \scriptsize 374,511 & \scriptsize 80.20 & \scriptsize 6.85 & \scriptsize  & \scriptsize  & \scriptsize  \\ \hline
\scriptsize Amazon (Tweb) & \scriptsize Miscellaneous & \scriptsize 403,364 & \scriptsize 12.10 & \scriptsize 2.81 & \scriptsize  & \scriptsize  & \scriptsize  \\ \hline
\scriptsize Dbpedia & \scriptsize Miscellaneous & \scriptsize 3,915,921 & \scriptsize 6.40 & \scriptsize 2.84 & \scriptsize  & \scriptsize  & \scriptsize  \\ \hline
\scriptsize Highland Tribes & \scriptsize Social & \scriptsize 16 & \scriptsize 7.20 & \scriptsize 1.22 & \scriptsize \textbf{0.73} & \scriptsize 0.76 & \scriptsize 2 \\ \hline
\scriptsize Sampson & \scriptsize Social & \scriptsize 18 & \scriptsize 14.00 & \scriptsize 1.00 & \scriptsize 0.28 & \scriptsize \textbf{0.28} & \scriptsize 2 \\ \hline
\scriptsize Taro Exchange & \scriptsize Social & \scriptsize 22 & \scriptsize 3.50 & \scriptsize 0.77 & \scriptsize \textbf{1.33} & \scriptsize 1.34 & \scriptsize 2 \\ \hline
\scriptsize Seventh Graders & \scriptsize Social & \scriptsize 29 & \scriptsize 17.20 & \scriptsize 1.56 & \scriptsize 0.43 & \scriptsize \textbf{0.43} & \scriptsize 3 \\ \hline
\scriptsize Dutch College & \scriptsize Social & \scriptsize 32 & \scriptsize 26.40 & \scriptsize 1.19 & \scriptsize 0.18 & \scriptsize \textbf{0.18} & \scriptsize 2 \\ \hline
\scriptsize Zachary Karate Club & \scriptsize Social & \scriptsize 34 & \scriptsize 4.60 & \scriptsize 2.32 & \scriptsize \textbf{1.18} & \scriptsize 1.33 & \scriptsize 2 \\ \hline
\scriptsize Highschool & \scriptsize Social & \scriptsize 70 & \scriptsize 7.80 & \scriptsize 2.61 & \scriptsize 1.37 & \scriptsize \textbf{1.36} & \scriptsize 3 \\ \hline
\scriptsize Physicians & \scriptsize Social & \scriptsize 117 & \scriptsize 7.90 & \scriptsize 2.22 & \scriptsize \textbf{1.60} & \scriptsize 1.61 & \scriptsize 2 \\ \hline
\scriptsize Residence Hall & \scriptsize Social & \scriptsize 217 & \scriptsize 16.90 & \scriptsize 2.66 & \scriptsize 1.48 & \scriptsize \textbf{1.48} & \scriptsize 2 \\ \hline
\scriptsize Adolescent Health & \scriptsize Social & \scriptsize 2,539 & \scriptsize 8.20 & \scriptsize 2.22 & \scriptsize 2.92 & \scriptsize \textbf{2.87} & \scriptsize 3 \\ \hline
\scriptsize Jung And Javax Dependency & \scriptsize Software & \scriptsize 6,120 & \scriptsize 16.40 & \scriptsize 3.75 & \scriptsize \textbf{2.14} & \scriptsize 2.44 & \scriptsize 9 \\ \hline
\scriptsize Jdk Dependency & \scriptsize Software & \scriptsize 6,434 & \scriptsize 16.70 & \scriptsize 3.81 & \scriptsize \textbf{2.17} & \scriptsize 2.47 & \scriptsize 8 \\ \hline
\scriptsize Linux & \scriptsize Software & \scriptsize 30,817 & \scriptsize 13.80 & \scriptsize 4.11 & \scriptsize \textbf{2.90} & \scriptsize 3.38 & \scriptsize 6 \\ \hline
\scriptsize Florida Ecosystem Dry & \scriptsize Trophic & \scriptsize 128 & \scriptsize 32.90 & \scriptsize 3.20 & \scriptsize 0.90 & \scriptsize \textbf{0.89} & \scriptsize 3 \\ \hline
\scriptsize Florida Ecosystem Wet & \scriptsize Trophic & \scriptsize 128 & \scriptsize 32.40 & \scriptsize 2.92 & \scriptsize 0.91 & \scriptsize \textbf{0.90} & \scriptsize 3 \\ \hline
\scriptsize Little Rock Lake & \scriptsize Trophic & \scriptsize 183 & \scriptsize 26.60 & \scriptsize 3.25 & \scriptsize 1.05 & \scriptsize \textbf{1.01} & \scriptsize 4 \\ \hline
\\\caption{Core-periphery statistics of KONECT networks. Number of nodes $N$, average degree $\langle k \rangle$, distance between two-block and $k$-core partitions VI (in bits), MDL per edge of hub-and-spoke model $\Sigma_{\mathcal{H}} / m$, MDL per edge of layered model $\Sigma_\mathcal{L} / m $, inferred number of layers $L$. MDLs are only reported for networks with less than 200,000 nodes (see Materials and Methods for details). All networks are drawn from KONECT, which is \href{http://konect.uni-koblenz.de/}{freely available online}.}
\label{appendix:tab:konect}
\end{longtable}
\endgroup

\clearpage

\subsection{Partition Similarity Comparison}

\begin{figure*}[!htb]
    \centering
    \includegraphics[scale=.42]{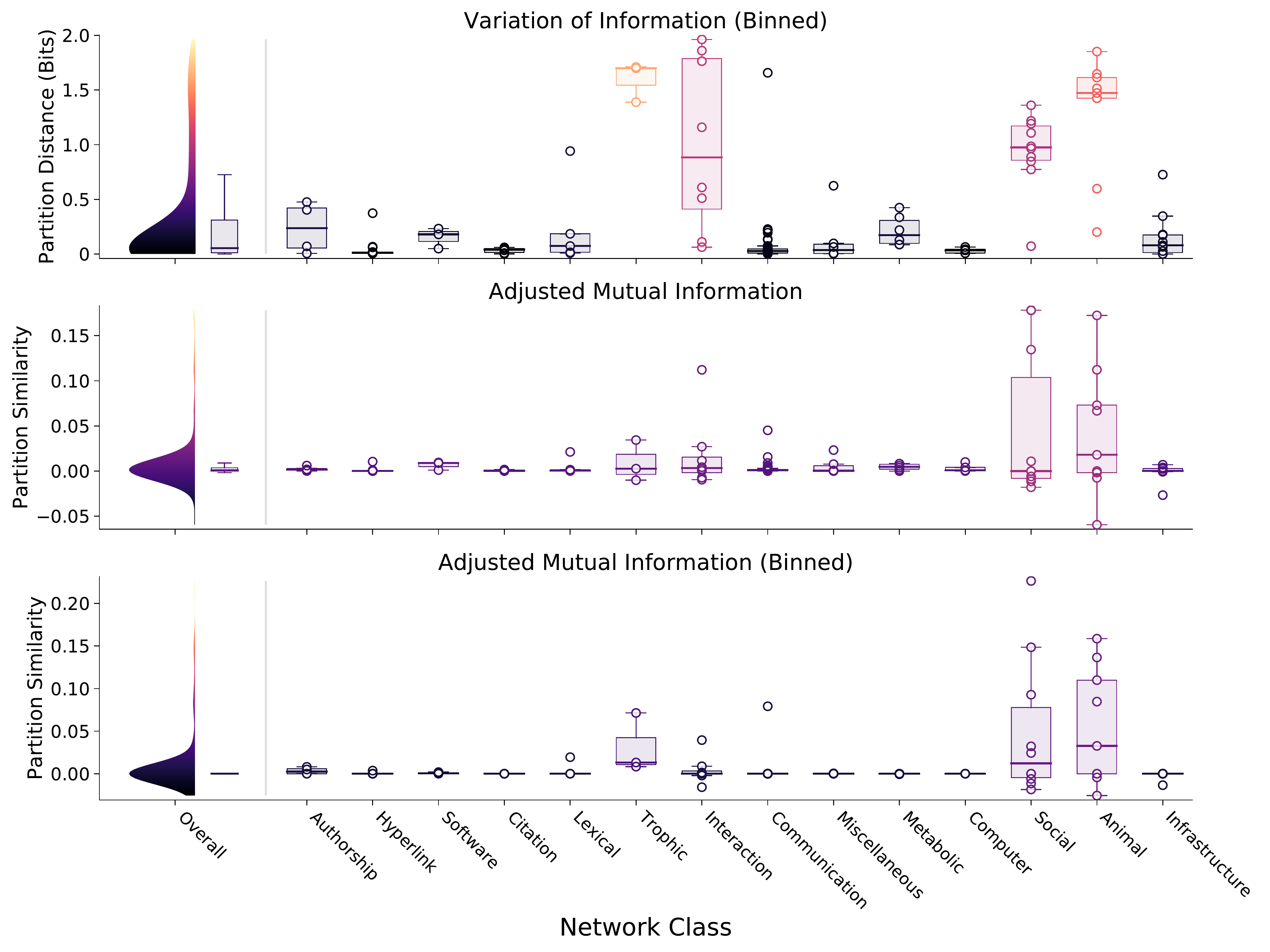}
    \caption{Similarity of two-block and $k$-cores partitions for different clustering similarity measures: variation of information (VI) \cite{meilua2007comparing} and adjusted mutual information (AMI) \cite{vinh2010information}. Note different vertical axis scales, and that VI is a distance measure (lower VI indicates more similar partitions) and AMI is a similarity measure (lower AMI indicates less similar partitions). ``Binned'' indicates similarity between the two-block partition and a binned $k$-cores partition, where the $k$-cores have been sequentially merged into two clusters close in size to those from the two-block partition. Network classes are sorted in the same order as Figure~\ref{fig01-similarities}, according to median VI within class.}
    \label{fig-sim-robustness}
\end{figure*}

\clearpage

\setcounter{figure}{0}
\section{Inference of Core-Periphery Stochastic Block Models}
\label{appendix-estimation}

We devise a Gibbs sampling procedure to generate samples from the joint posterior distribution $P(\theta,\p \mid \A)$ of the core-periphery stochastic block models. The general idea is to alternate between sampling the densities $\p$ conditioned on the block assignments $\theta$, and sampling $\theta$ conditioned on $\p$. This method provably generates samples from the target joint posterior distribution under very general conditions. We use the samples from the joint distribution $P(\theta,\p \mid \A)$ to construct the estimated partition $\hat{\theta}$ by marginalization \cite{decelle2011asymptotic}, i.e., by setting $\hat{\theta}_i=r$ where $r$ is the block in which node $i$ is most often found among samples.

In what follows, for simplicity of notation, we derive the Gibbs sampler for the layered model, but the procedure we present here immediately maps onto the case of the hub-and-spoke model.

\subsection{Gibbs and rejection sampling for $P(\p \mid \A, \theta)$}

Let $\alpha_s = \sum_{r \leq s} m_{rs}$ and $\beta_s = \sum_{r \leq s} M_{rs} - m_{rs}$, where $m_{rs}$ and $M_{rs}$ are the number of edges that do exist and that could potentially exist between blocks $r$ and $s$ respectively. That is, $M_{rs} = n_r n_s$ if $r \not = s$ and $M_{rs} = n_r (n_r - 1) / 2$ if $r = s$. Briefly note that by rearranging the likelihood presented in the \hyperlink{methods-block-models}{Materials and Methods} such that it takes the product over layers rather than blocks, we can write the posterior of the layered model as
\begin{equation}
    P(\theta, \p \mid \A)
        \propto 
            \prod_{s=1}^{\ell} p_s^{\alpha_s} (1 - p_s)^{\beta_s}
            \cdot \mathds{1}_{\{0 < p_\ell < \ldots < p_1 < 1\}}.
\end{equation}

Now, for notation define $\p_{-s} = (p_1, \ldots, p_{s-1}, p_{s+1}, \ldots, p_{\ell})$. First observe that 

\begin{equation}
P(\p \mid \A, \theta) 
     = 
        \frac{P(\theta, \p \mid \A)}{P(\theta \mid \A)}
    \propto
        P(\A \mid \theta, \p)P(\p),
\end{equation}
and, furthermore, that
\begin{equation}
P(\p \mid \A, \theta)
    =
        P(p_1, \ldots, p_\ell \mid \A, \theta)
    =
        P(p_s \mid \A, \theta, \p_{-s})
        P(\p_{-s} \mid \A, \theta).
\end{equation}
Hence, rearranging terms, we have
\begin{equation}
    P(p_s \mid \A, \theta, \p_{-s})
        \propto
            P(\A \mid \p, \theta)P(\p)
\end{equation}
where the terms absorbed by proportionality do not depend on $p_s$. Therefore,
\begin{equation}
    P(p_s \mid \A, \theta, \p_{-s})
        \propto
                p_s^{\alpha_s} (1 - p_s)^{\beta_s} 
                    \cdot 
                    \mathds{1}_{\{0 < p_\ell < \ldots < p_1 < 1\}}.
\end{equation}
In other words, each $p_s$ conditioned on the other model parameters follows a truncated beta distribution $\text{Beta}(\alpha_s + 1, \beta_s + 1)$. 

For a fixed partition $\theta$, drawing a series of densities from $P(\p \mid \A, \theta)$ can therefore be done by Gibbs sampling, by drawing new $p_s^{(t+1)}$ sequentially, from $s=1$ to $s=\ell$. At Gibbs sample step $t+1$, we have to draw from a truncated beta distribution such that $p_{s - 1}^{(t+1)} < p_s^{(t+1)} < p_{s + 1}^{(t)}$.  We construct a rejection sampling procedure that takes advantage of the fact that we know the parametric form of the draws for $p_s^{(t+1)}$. We break the sampling down into two conditions, based on the location of the peak $\mu^{(t)}_s = (\alpha_s^{(t)} + 1) / (\alpha_s^{(t)}  + \alpha_s^{(t)} + 2)$, where $\alpha_s^{(t)}$ and $\beta_s^{(t)}$ are calculated with the known (fixed partition). 

\begin{enumerate}
    \item If $p_{s-1}^{(t+1)} < \mu^{(t)}_s < p_{s+1}^{(t)}$, then the peak is within the truncation constraints, then we can make a naive draw of $p_s^{(t+1)}$ from the beta distribution and reject until we obtain a draw that respects the constraint.\\
    \item If $\mu^{(t)}_s < p_{s-1}^{(t+1)}$ or $\mu^{(t)}_s > p_{s+1}^{(t)}$, then the peak is outside the truncation constraints and drawing from an unconstrained beta distribution will not work very well. We instead use a rejection sampling approach. Let $f$ be the functional form of the unconstrained Beta distribution appearing in our model:
    \begin{equation}
        f(x; \alpha_s^{(t)}, \beta_s^{(t)}) 
            = 
                \frac{x^{\alpha_s^{(t)}} (1 - x)^{\beta_s^{(t)} }}{B \bigl(\alpha_s^{(t)} + 1, \beta_s^{(t)} + 1 \bigr)},
    \end{equation}
    where $B(a,b)=\Gamma(a)\Gamma(b)/\Gamma(a+b)$ and $\Gamma(x)$ is the Gamma function.
    Next, let $g$ be the functional form of the uniform distribution over $[p_{s-1}^{(t+1)}, p_{s+1}^{(t)}]$,
    \begin{equation}
        g(x) = \frac{1}{p_{s+1}^{(t)} - p_{s-1}^{(t+1)}}.
    \end{equation}
    Finally, choose
    \begin{equation}
        M
            =
                \left(p_{s+1}^{(t)} - p_{s-1}^{(t+1)}\right)
                \cdot \max\biggl(f\bigl(p_{s-1}^{(t+1)}\bigr), f\bigl(p_{s+1}^{(t)} \bigr)\biggr).
    \end{equation}
    Then, by this choice of $M$, $f(x) < M g(x)$ over the interval $[p_{s-1}^{(t+1)}, p_{s+1}^{(t)}]$. Therefore, using the rejection sampling technique, if we draw a sample $u$ uniformly over that interval and accept it with probability $f(u) / M g(u)$, we will sample from the truncated beta distribution appropriately. 
\end{enumerate}

This rejection sampling procedure is efficient because the distribution is strongly peaked. To see this, observe that the the variance $\sigma_s^2$ of the beta distribution $\text{Beta}(\alpha_s + 1, \beta_s + 1)$ is
\begin{equation}
    \sigma_s^2 = \frac
            { \left( \sum_{r \leq s} m_{rs} \right) \left( \sum_{r \leq s} M_{rs} - m_{rs} \right)}
            { \left( \sum_{r \leq s} M_{rs} \right)^2 \left( 1+ \sum_{r \leq s} M_{rs} \right)^2}
\end{equation}
Note that the denominator of the variance $\sigma_s^2$ grows with the number of potential links extending from the layer as $O\bigl( (\sum_{r \leq s} M_{rs} )^4 \bigr)$. Further, recall that the number of potential links within the network as a whole grows as $O(N^2)$. Consequently, the beta distribution $B(\alpha_s, \beta_s)$ from which we draw $p^{(t+1)}_s$ is very narrowly peaked right around the mean of the distribution $\mu_s^{(t)}$.

% -------------------------------

\subsection{Sampling $P(\theta \mid \A, \p)$ with Markov Chain Monte Carlo}

The next step in the Gibbs sampler requires us to draw from $P(\theta \mid \p, \A) \propto P(\A \mid \theta, \p) P(\theta)$. To do this, we apply a label-switching Markov chain Monte Carlo (MCMC) procedure \cite{peixoto2017bayesian,peixoto2014efficient}. Specifically, we first initialize each node with a random block label. For a fixed number of iterations, we then choose a random node $i$ and switch its block label. With the new partition $\theta'$ created by flipping $\theta_i$, we compute the Metropolis-Hastings criterion probability $a$,
\begin{equation}
    a
        =
            \min\left(1, \frac{P(\theta' \mid \p, \A)}{P(\theta \mid \p, \A)} \frac{P(\theta \mid \theta')}{P(\theta' \mid \theta)} \right),
\end{equation}
which prescribes the probability of accepting the proposed move of $\theta$ to $\theta'$. The simplest algorithm for proposing a move for $i$ is to choose a new label completely at random, and so
\begin{equation}
    P(\theta'_i \mid \theta)
        =
            \frac{1}{\ell}.
\end{equation}
This updating schema preserves ergodicity for the MCMC procedure.  Another, generally more efficient, strategy has also been proposed in the literature which empirically often leads to less rejections of proposed moves in the parameter space \cite{peixoto2017bayesian,peixoto2014efficient}. The proposal strategy leverages information from the node's neighborhood and proposes a new label $r$ with probability
\begin{equation}
    P(\theta_i = r \mid \theta)
        =
            \sum_s \sum_j 
                \frac{A_{ij} \delta_{\theta_j,s}}{k_i} 
                \frac{m_{sr} + \varepsilon}{m_s + \varepsilon \, \ell},
\end{equation}
where $m_s$ is the total number of edges connecting to block $s$ and $\varepsilon$ is a relatively arbitrary parameter for preserving ergodicity that can be set to be small. We applied both label-switching proposals and found mixed results. For the discernment experiment (see Figure~\ref{fig-synthetic-networks}A), there was no difference in the results when using either random proposals or neighborhood proposals. For the number of layers experiment (see Figure~\ref{fig-synthetic-networks}B), we found that the neighborhood proposals led to MCMC chains that did not converge as well to the ground truth partition over the course of the Gibbs sampling, as compared to the random proposals. For the reported results here, we used uniformly random label proposals.

% -------------------------------

\subsection{Algorithmic Complexity and Implementation}

Drawing from $P(\p \mid \theta, \A)$ and evaluating the Metropolis-Hastings criterion when sampling $P(\theta \mid \p, \A)$ both require knowing the number of edges within and between blocks. The number of edges between blocks $m_{rs}$ can be calculated once in $O(M)$ time and then stored in memory. If the number of nodes in each block is also recorded, then the number of potential edges $M_{rs}$ can be updated in $O(\ell^2)$ time, where $\ell$ is the number of blocks (which is always $\ell = 2$ for the hub-and-spoke model). Similarly, the evaluation of the posterior involving the product (sum in log space) over the inter-block connectivities $r < s$ can be done in $O(\ell^2)$ time. Let $T_{\text{Gibbs}}$ be the number of Gibbs sampling iterations and let $T_{\text{MCMC}}$ be the number of iterations to run the MCMC chain within each iteration of the Gibbs sampling. Then, the overall time complexity of the core-periphery inference algorithm is $O(M + \ell^2 \, T_{\text{Gibbs}} \, T_{\text{MCMC}})$. 

We summarize the entire inference procedure in Algorithm~\ref{core-periphery-alg} and provide Python code implementing both the hub-and-spoke and layered core-periphery models at \url{https://github.com/ryanjgallagher/core_periphery_sbm}.

\begin{algorithm}[H]
\caption{Core-Periphery SBM Inference}
    \begin{algorithmic}
    \Procedure{CorePeriphery}{$G$, $T_{\text{Gibbs}}$, $T_{\text{MCMC}}$}
        % -------------------------- Initialize algorithm --------------------------
        \State $\theta^{(0)} \leftarrow$ Draw random block assignments \Comment{Initialize inference}
        \State $\n^{(0)},\m^{(0)},\M^{(0)} \leftarrow$ CalculateBlockStatistics$\bigl(\theta^{(0)}\bigr)$
        \State $\theta^{(0)},\n^{(0)},\m^{(0)},\M^{(0)} \leftarrow$ Reorder blocks to obey density constraints
        \For{$p_i \in \p$}
            \State $p^{(0)}_i \leftarrow$ SampleTruncBeta$\bigl(\m^{(0)}, \M^{(0)}, \p \bigr)$
        \EndFor
        \newline
        
        % -------------------------- Gibbs Sample --------------------------
        \For{$t \leftarrow 1, T_{\text{Gibbs}}$}
        % -------------------------- Sample labels --------------------------
            \State $\theta^{(t)} \leftarrow \theta^{(t-1)}$ \Comment{Initialize Gibbs iteration}
            \State $\n^{(t)}, \m^{(t)}, \M^{(t)} \leftarrow$ 
                CalculateBlockStatistics$\bigl(\theta^{(t)}\bigr)$
                \newline
            
            \For{$\tau \leftarrow 1, T_{\text{MCMC}}$} \Comment{Sample $P(\theta \mid \A, \p^{(t-1)})$}
                \State $i \leftarrow$ Draw random node
                \State $r \leftarrow$ Draw random block
                \State $\theta' \leftarrow \theta^{(t)}$,  
                \State $\theta'_i \leftarrow r$
                \State $\n', \mathbf{m}', \mathbf{M}' \leftarrow$ 
                    CalculateBlockStatistics$\bigl(\theta' \bigr)$
                \State $a \leftarrow$ 
                    MetropolisHastings$\bigl(P(\theta^{(t)} \mid \A, \p), \, P(\theta' \mid \A, \p)\bigr)$
                \If{$a = 1$}
                    \State $\theta^{(t)},\n^{(t)},\m^{(t)},\M^{(t)} \leftarrow
                            \theta',\n',\m',\M'$
                \EndIf
            \EndFor
            \newline
            
            % -------------------------- Sample ps --------------------------
            \For{$p_i \in \p$} \Comment{Sample $P(\p \mid \A, \theta^{(t)})$}
                \State $p^{(t)}_i \leftarrow$ SampleTruncBeta$\bigl(\m^{(t)}, \M^{(t)}, \p \bigr)$
            \EndFor
            \newline
        \EndFor
    \EndProcedure
    \end{algorithmic}
\label{core-periphery-alg}
\end{algorithm}

\clearpage

\setcounter{figure}{0}
\section{Calculating Model Fit via Description Length}
\label{appendix-model-fit}

\subsection{Approximating Description Length}

Recall that we let $\theta_\mathcal{H}$ be a partition inferred by the hub-and-spoke model $\mathcal{H}$, and similarly we let $\theta_\mathcal{L}$ be a partition inferred by the layered model $\mathcal{L}$. We consider the posterior odds ratio,
\begin{equation}
    \Lambda
        =
            \frac{
                P(\theta_\mathcal{H}, \mathcal{H} \mid \A)
            }{
                P(\theta_\mathcal{L}, \mathcal{L} \mid \A)
            }
        = 
            \frac{
                P(\A, \theta_\mathcal{H} \mid \mathcal{H}) P(\mathcal{H})
            }{
                P(\A, \theta_\mathcal{L} \mid \mathcal{L}) P(\mathcal{L})
            },
\end{equation}
which measures which partition is more likely given our observed network data. Briefly note that for both models we have removed any reference to the parameters $\p$. This is by design: we are primarily interested in evaluating the final \emph{partition} of nodes, and we do not want our model comparison to be sensitive to $\p$, which may fluctuate unstably over real numbers. If we assume that the models are equally likely with $P(\mathcal{H}) = P(\mathcal{L}) = 1/2$, then 
\begin{equation}
    \Lambda
        =
            \frac{
                P(\A, \theta_\mathcal{H} \mid \mathcal{H})
            }{
                P(\A, \theta_\mathcal{L} \mid \mathcal{L})
            }.
\end{equation}
This implies that
\begin{equation}
    - \Lambda
        = 
            \underbrace{
                    -\log P(\A, \theta_\mathcal{H} \mid \mathcal{H})
                }_{
                    \Sigma_\mathcal{H}
                }
            - \biggl(
            \underbrace{
                    -\log P(\A, \theta_\mathcal{L} \mid \mathcal{L})
                }_{
                    \Sigma_\mathcal{L}
                }
            \biggr),
\end{equation}
where $\Sigma_\mathcal{M}$ is the description length of model $\mathcal{M}$. Consistent with the posterior odds ratio, the difference in description lengths, $\Sigma_\mathcal{H} - \Sigma_\mathcal{L}$, is negative if the hub-and-spoke model is a better fit than layered model, and positive otherwise. That is, the model with the minimum description length between the two is the best descriptor of a network's structure. To calculate model fit, then, it suffices to be able to calculate the description length $\Sigma_\mathcal{M}$ for each model $\mathcal{M}$.

Unfortunately, the description length is given by an integral with (what appears to be) no analytical closed form for both core-periphery block models:
\begin{align}
    \label{eq:dl_generic}
    P(\A, \theta_\mathcal{M} \mid \mathcal{M})
        &=
            \int
                P(\A, \theta_\mathcal{M}, \p \mid \mathcal{M})
                \, d\p \\
        &= \int
                P(\A \mid \theta_\mathcal{M}, \p, \mathcal{M})
                P(\theta_\mathcal{M} \mid \mathcal{M})
                P(\p \mid \mathcal{M})
                \, d\p.
\end{align}
The constraints on $\p$ and the likelihood leads to complicated polynomials integrated over a simplex.
Hence we have to use numerical methods to approximate the integral.
Due to the high dimensionality of $\p$, we resort to a Monte-Carlo approximation.
The most straightforward way to perform Monte-Carlo is to draw $n$ samples from the prior $P(\p)$, leading to
\begin{equation}
    P(\A, \theta_\mathcal{M} \mid \mathcal{M})
        \approx
            \frac{P(\theta_\mathcal{M})}{n}
            \sum_{i = 1}^n
                P(\A \mid \theta_\mathcal{M}, \p_i, \mathcal{M}).
\end{equation}
The block connectivity prior $P(\p)$ specifies a uniform distribution over all $\p$ that satisfy an ordering constraint $0 < p_{\ell} < p_{\ell-1} < \ldots < p_{1} < 1$. (The constraint is almost identical for both the hub-and-spoke model.)
An efficient way to draw a direct sample from $P(\p)$ is to draw the \emph{spacings} $p_{r-1}-p_r$ between two ordered $p_{r} < p_{r-1}$. These spacings sum to one and all combinations of spacings are equally likely. Therefore, we may draw all of them simultaneously by generating a normalized vector $\bm{\pi}$ of length $\ell + 1$ from
\begin{equation}
    \mathbf{\pi}
        \sim
            \text{Dirichlet}(\mathbf{1}_{\ell + 1}),
\end{equation}
where $\mathbf{1}_{\ell + 1}$ is a vector of length $\ell + 1$ consisting entirely of ones.
We then simply have to define
\begin{equation}
    p_s
        =
            1 - \sum_{r \leq s} \pi_s.
\end{equation}

To evaluate the description length $\Sigma_\mathcal{M}$ numerically and robustly, one can then use the log-sum exponent trick
\begin{equation}
    \log P(\A, \theta_\mathcal{M} \mid \mathcal{M})
        \approx
            L_\text{max}
            - \log n
            + \log P(\theta_\mathcal{M} \mid \mathcal{M})
            + \log \sum_{i = 1}^n
                e^{\Delta(p_i)},
\end{equation}
where
\begin{equation}
    L_\text{max}
        =
            \max_i \{ \log P(\A \mid \theta_\mathcal{M}, \p_i, \mathcal{M}) \}
\end{equation}
is the empirical maximum over log likelihoods, and
\begin{equation}
    \Delta(\p_i)
        =
            \log P(\A \mid \theta_\mathcal{M}, \p_i, \mathcal{M})
            -
            L_\text{max}
\end{equation}
is the deviation from that maximum for sample $i$. All together, this method then yields an estimate of the difference in description lengths, $\Sigma_\mathcal{M} - \Sigma_\mathcal{L}$.

% -----------------------------

\subsection{Improving Approximations through Importance Sampling}

Some problems may arise if $P(\p)$ does not have much probability mass where $P(\A \mid \theta_\mathcal{M}, \p, \mathcal{M})$ is the largest. This problem can become particularly pronounced when the number of layers $\ell$ is large in the layered model.
Hence we also introduce a reweighing technique --- importance sampling --- that can enhance convergence in these problematic cases.
We first rewrite the integral appearing in Eq.~\eqref{eq:dl_generic} as
\begin{align}
    P(\A, \theta_\mathcal{M} \mid \mathcal{M})
        = \int
            P(\A \mid \theta_\mathcal{M}, \p, \mathcal{M})
            P(\theta_\mathcal{M} \mid \mathcal{M})
            \frac{
                    P(\p \mid \mathcal{M})
                }{
                    Q(\p \mid \mathcal{M})
                }
            Q(\p \mid \mathcal{M})
            \, d\p 
\end{align}
where $Q(\p \mid \mathcal{M})$ is some distribution of our choosing, called the proposal distribution.
We can then use Monte-Carlo sampling again and evaluate the description length as
\begin{align}
    P(\A, \theta_\mathcal{M} \mid \mathcal{M})
            \approx
            \frac{P(\theta_\mathcal{M})}{n}
            \sum_{i = 1}^n
                P(\A \mid \theta_\mathcal{M}, \p_i, \mathcal{M})
                \frac{
                        P(\p_i \mid \mathcal{M})
                    }{
                        Q(\p_i \mid \mathcal{M})
                    },
\end{align} 
To bias the approximation towards values of $\p_i$ that have more mass according to the likelihood $P(\A \mid \theta_\mathcal{M}, \p_i, \mathcal{M})$, we choose to write
\begin{align}
    Q(\p \mid \mathcal{L})
        &=
            Q(p_1, p_2, \ldots, p_\ell) \\
        &=
            Q(p_\ell \mid p_1, \ldots, p_{\ell - 1})
                \,
                Q(p_{\ell-1} \mid p_1, \ldots p_{\ell - 2})
                \ldots
                Q(p_2 \mid p_1)
                \,
                Q(p_1) \\
        &=
            Q(p_\ell \mid p_{\ell - 1}) 
                \ldots 
                Q(p_2 \mid p_1)
                Q(p_1),
\end{align}
for the layered model (and similarly for the hub-and-spoke model), where the last equality is valid given our core-periphery constraint $p_1 > \ldots > p_\ell$. We then specify $Q$ by choosing
\begin{align}
    p_1
        &\sim 
            \text{Uniform}(0, 1)\\
    p_{r+1} \mid p_r
        &\sim
            \text{Uniform}(0, p_r).
\end{align}
This recursive distribution, which satisfies the core-periphery constraint, geometrically spaces $\p$ between $[0, 1]$ on average, biasing the draws to have lower values of $p_r$, as in real networks. Because we know the functional form of the proposal distribution $Q(\p \mid \mathcal{M})$ and the original prior $P(\p \mid \mathcal{M})$, we can reweight our approximation accordingly, giving us an estimate of the description length which is more stable for model selection.

\end{document}